\documentclass[
aps,
prc,
superscriptaddress,
floatfix,
nofootinbib,
twocolumn
]{revtex4-2}

\usepackage{amsmath,amssymb,amsfonts,physics,graphicx,float}

\usepackage[setpagesize=false]{hyperref}
\allowdisplaybreaks[4]

\makeatletter
\g@addto@macro\appendix{%
  \setcounter{figure}{0}%
  \setcounter{table}{0}%
  \@addtoreset{figure}{section}%
  \@addtoreset{table}{section}%
}
\makeatother

\usepackage{ulem}
\usepackage{color}
\definecolor{ar}{rgb}{1.0, 0.01, 0.24}
\definecolor{al}{rgb}{0.82, 0.1, 0.26}
\definecolor{ev}{rgb}{0.56, 0.0, 1.0}

\usepackage{bm}

\begin{document}

\title{Evidence for a Continuous Hadron–Quark Transition in Cold Dense Matter}

\author{Yong-Jia Huang}
\email{huangyj@pmo.ac.cn}
\affiliation{Key Laboratory of Dark Matter and Space Astronomy,
Purple Mountain Observatory, Chinese Academy of Science, Nanjing, 210023, China}
\affiliation{RIKEN Center for Interdisciplinary Theoretical and Mathematical Sciences (iTHEMS), RIKEN, Wako 351-0198, Japan}

\author{Bikai Gao}
\email{bikai@rcnp.osaka-u.ac.jp}
\affiliation{Research Center for Nuclear Physics, The University of Osaka, Ibaraki, Osaka 567-0047, Japan}

\date{\today}

\begin{abstract}
Statistical evidence for a continuous hadron-quark transition is found in this work. Confronting microscopic descriptions connecting the Parity Doublet Model and the Nambu--Jona-Lasinio model with observationally constrained non-parametric equations of state, Bayesian model comparison decisively favors a boundary-free crossover over the conventional first-order Maxwell construction ($\Delta\ln\mathcal{Z} \sim +7.20$) and fixed-boundary crossover ($\Delta\ln\mathcal{Z} \sim +5.16 $). The preferred crossover decouples the intermediate stiffening from the intrinsic stiffness of each phase, allowing both sectors to exhibit physical self-consistency. 
The chiral-invariant nucleon mass is large, \(M_0 = 834^{+28}_{-92}\ \text{MeV}\), as expected for a substantial baryon mass surviving chiral restoration. While the quark sector accommodates a small pairing gap $\Delta_{\rm CFL} = 88^{+83}_{-59}\text{ MeV}$ consistent with perturbative QCD limits, avoiding the excessively large gaps $\Delta_{\rm CFL} \gtrsim 200\text{ MeV}$ required by the other constructions. These results demonstrate that a realistic unified description represents a continuous transition deviating substantially from isolated effective models of hadrons and quarks. Exploring genuine phase boundaries is therefore necessary through the emergence of spinodal instabilities within unified frameworks.
\end{abstract}

\maketitle

%\tableofcontents

%%%%%%%%%%%%%%%%%%%%%%%%%%%%%%%%%%%%%%%%%%%%%%%%%%%%%%%%%%%%%%%%
\section{Introduction}
\begin{figure}[htbp]
\centering
\includegraphics[width=\linewidth]{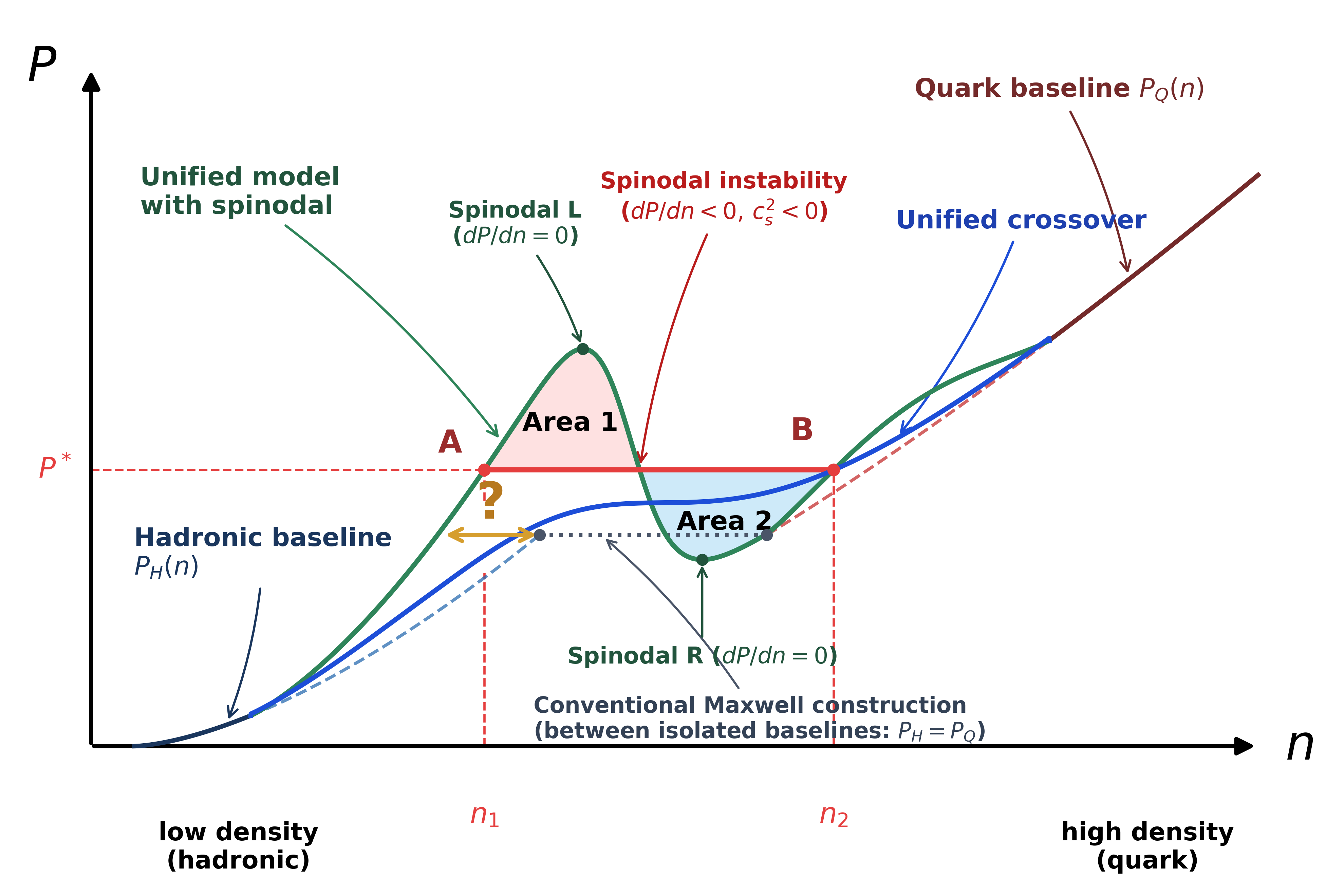}
\caption{Schematic pressure--density ($P$--$n$) relations comparing different transition paradigms: (1) a unified model with a first-order phase transition (green curve) exhibiting a spinodal loop ($dP/dn < 0$) resolved by the Maxwell equal-area construction (${\rm Area~1} = {\rm Area~2}$ at $P^*$ between points A and B); (2) a conventional two-phase Maxwell construction connecting isolated hadronic ($P_H$) and quark ($P_Q$) models (dotted horizontal plateau, with the mismatch to the unified model indicated by the question mark); (3) a unified crossover (blue curve) which continuously connects the hadronic and quark regimes without a sharp phase boundary or thermodynamic instability.}
\label{fig:schematic}
\end{figure}

The state of cold dense matter in Quantum Chromodynamics (QCD) remains a central problem in nuclear physics~\cite{Fukushima:2013rx,Baym:2017whm}. At zero baryon chemical potential, lattice QCD and heavy-ion collision experiments establish a smooth crossover~\cite{Bazavov:2014pvz}. In contrast, at zero temperature and supranuclear densities, direct first-principles calculations remain unavailable due to the fermion sign problem. Investigating the nature of matter in this cold, dense regime therefore requires effective field theories that faithfully embody these non-perturbative features in each sector. For hadronic matter, an effective description is provided by the parity doublet model (PDM)~\cite{Motohiro:2015taa,Gao:2024chh}, which treats $N(939)$ and $N(1535)$ as chiral partners and allows both masses to approach a nonzero chiral-invariant mass ($M_0$) when the chiral condensate vanishes~\cite{Detar:1988kn,Jido:2001nt,Motohiro:2015taa,Marczenko:2020jma,Gao:2025nkg,Kawaguchi:2025cuf,Gao:2026scv,Kunihiro:2026gjo}. This mass structure is qualitatively supported by finite-temperature lattice calculations, where negative-parity baryon masses decrease toward those of their positive-parity partners near the chiral crossover, while the positive-parity masses almost do not change~\cite{Aarts:2015mma,Aarts:2017rrl,Aarts:2018glk}. It is further substantiated by QCD sum-rule analyses predicting a substantial nucleon mass in a chiral symmetric vacuum~\cite{Kim:2020zae,Lee:2023ofg}, and more generally by the QCD trace anomaly, which generates a dynamical mass scale even when current-quark masses vanish~\cite{Ji:1994av}. For deconfined quark matter, the three-flavor Nambu--Jona-Lasinio (NJL)-type model~\cite{Baym:2017whm,Li:2019ztm,Xia:2024wpz} provides an effective microscopic description by self-consistently capturing dynamical chiral symmetry restoration, vector repulsion essential for stabilizing dense matter, and color-flavor locked (CFL) diquark pairing. Connecting these regimes usually relies on the Maxwell construction between individually stable hadronic and quark equations of state (EOS) (See Fig.~\ref{fig:schematic}). However, whether such conventional two-phase constructions capture the actual thermodynamics of dense matter remains an open question. 

From the perspective of statistical mechanics, an intrinsic first-order
phase transition does not originate from two disconnected physical
systems. As established by Lee and Yang~\cite{Yang:1952be,Lee:1952pf},
coexisting phases 
are described by a single grand partition function in the thermodynamic limit. In a mean-field
description, a first-order transition can exhibit a non-monotonic
van der Waals loop with a unstable spinodal branch
($dP/dn < 0$ at T=0)
~\cite{Lebowitz:1966,Chomaz:2003dz,Randrup:2009gp}, where the Maxwell
construction replaces this loop with a coexistence plateau.
Although the hadron--quark transition in cold dense matter is microscopically more complex than ordinary
liquid--gas phase transitions and may involve an interplay
between chiral dynamics,
color superconductivity, and quark
delocalization as hadrons overlap~\cite{Alford:2007xm,Baym:2017whm,McLerran:2018hbz},
their basic thermodynamic principles remain the same. Theoretically, the 
possible interplay among these microscopic processes motivates a realistic unified description
of dense matter. This raises the fundamental question of whether the thermodynamic behavior of dense matter in the
transition region  deviates significantly
from the predictions of separate hadronic and quark effective models.

Recent multi-messenger observations of neutron stars (NSs) now make it possible to address this question quantitatively. Combining observational data from NICER~\cite{Miller:2019cac,Riley:2019yda,Salmi:2024aum,Choudhury:2024xbk,Mauviard:2025dmd}, gravitational waves (GW170817)~\cite{LIGOScientific:2017ync,LIGOScientific:2018cki}, and theoretical limits~\cite{Drischler:2020hwi,Komoltsev:2021jzg} within a non-parametric framework, the various information could be summarized as the thermodynamically consistent statistical EOS posterior \cite{Huang:2026jgl}. This non-parametric reconstruction favors a characteristic sound-speed profile featuring a prominent peak exceeding the conformal limit ($c_s^2 > 1/3$) followed by extended softening with $c_s^2 < 1/3$ over a broad density range.

The physical origin of this non-monotonic structure is naturally provided by the hadron-quark crossover picture (see also Refs.~\cite{Dexheimer:2009hi,Horowitz:1985gv,Ropke:1986qs} for other kinds of unified models). Since composite nucleons have a finite spatial size, increasing density can cause them to overlap geometrically, potentially forming an interconnected percolation network and motivating a crossover description~\cite{Baym:1979,Asakawa:1995zu,Baym:2017whm}. The associated quark Pauli blocking can stiffen the matter before full deconfinement is achieved~\cite{Oka:1980ax,Kojo:2015fua,Kojo:2020krb}. Thermodynamically, this crossover picture can be phenomenologically realized through two different formulations with the equal number of transition parameters: one parameterizes the boundary densities where matter leaves the pure hadronic regime and enters the pure quark regime, with the intermediate EOS analytically determined by boundary matching conditions~\cite{Baym:2019iky,Kojo:2021wax,Geng:2026hbf};the other formulates a boundary-free energy-density interpolation $\varepsilon(n) = [1-w(n)]\varepsilon_{\rm H}(n) + w(n)\varepsilon_{\rm Q}(n)$ governed by phenomenological crossover parameters, where thermodynamic consistency requires an additional term $n w'(n)[\varepsilon_{\rm Q}(n) - \varepsilon_{\rm H}(n)]$ that can enlarge the pressure and sound speed in the crossover region~\cite{Masuda:2012kf,Masuda:2012ed}.

In this Letter, we use the statistical EOS posterior as a reconstructed thermodynamic constraint to perform a Bayesian model comparison among phenomenological transition scenarios connecting the PDM and the three-flavor NJL-type quark model. To summarize the information from both microscopic calculations and integrated macroscopic observables, we confront candidate equations of state directly in correlated $(P, c_s^2)$ space across three transition constructions: a conventional Maxwell construction, a fixed-boundary polynomial crossover~\cite{Baym:2019iky,Kojo:2021wax}, and a boundary-free energy-density crossover~\cite{Masuda:2012kf}. Within the adopted model set and priors, our Bayesian analysis reveals decisive statistical evidence favoring a continuous crossover over the conventional Maxwell construction, with the boundary-free formulation strongly preferred. Within the favored crossover, the inferred parameters in both sectors naturally achieve physical self-consistency, demonstrating that multi-messenger observations favor a continuous hadron–quark transition whose thermodynamic behavior differs from extrapolations of the separate hadronic and quark effective models.

\section{Method}
\label{sec:method}
We evaluate microscopic EOS by confronting them with the non-parametric~\cite{Landry:2018dwm,Essick:2019gia} statistical EOS posterior obtained in our companion work~\cite{Huang:2026jgl}, which synthesizes constraints from chiral effective field theory, multi-messenger observations from NICER, GW170817, and massive pulsars, alongside perturbative QCD limits into $\sim 10^4$ weighted samples. To construct a continuous likelihood function, each sample is discretized on a 50-point logarithmic grid spanning $n_B \in [0.16,\,1.40]\text{ fm}^{-3}$ ($[n_0,\,8.75\,n_0]$). The concatenated values of pressure $P(n_B)$ and sound speed squared $c_s^2(n_B)$ form a 100-dimensional representation for each EOS. A weighted Principal Component Analysis (PCA)~\cite{Jolliffe2002} reduces this dimensionality, and retaining the first 10 components captures over 95\% of the cumulative variance. Fitting a weighted Gaussian kernel density estimator in this 10-dimensional PCA space yields a continuous likelihood functional $\mathcal{L}(\mathrm{EOS})$ preserving the correlated joint distribution between pressure and sound speed across the sampled densities.

To explore the transition mechanism, we construct three representative transition frameworks sharing identical baseline hadronic and quark models. The hadronic sector is described by the PDM parametrized by $(M_0,\,L_0)$, and the quark sector is modeled by the three-flavor NJL-type quark model with color superconductivity parametrized by $(H/G,\,g_V/G,\,B)$, forming a shared baseline vector $\bm{\phi} = (M_0,\,L_0,\,H/G,\,g_V/G,\,B)$. We adopt uniform priors on all shared parameters: $M_0 \in [600, 900]\text{ MeV}$, $L_0 \in [40, 70]\text{ MeV}$, $H/G \in [0.4, 2.0]$, $g_V/G \in [0.05, 1.5]$, and the bag constant $B \in [0, 300]\text{ MeV/fm}^3$. We connect these sectors via three schemes, namely (i) a first-order Maxwell construction with the transition chemical potential $\mu_c$ determined by $P_H(\mu_c) = P_Q(\mu_c)$ and no additional parameters are needed, (ii) a fixed-boundary crossover smoothly matching thermodynamic potentials within a window $\bm{\psi} = (n_H,\,n_Q)$ using a fifth-order polynomial (Poly5), and (iii) a boundary-free crossover with connection parameters $\bm{\psi} = (\bar{n},\,\Gamma)$, where thermodynamic consistency naturally introduces an emergent term $n w'(n)[\varepsilon_Q - \varepsilon_H]$ that enhances the pressure and sound speed. Full microscopic formulations are documented in Appendix~\ref{sec-eos}.

To compare the three transition patterns in a common $\bm{\phi}$ space, we treat $\bm{\psi}$ as nuisance parameters and evaluate $\mathcal{L}_{\mathrm{eff}}(\bm{\phi}) =\int\mathcal{L}(\bm{\phi},\bm{\psi})\pi(\bm{\psi})\,d\bm{\psi}$, where $\pi(\bm{\psi})$ is uniform on the connection parameter domain ($n_H \in [0.24, 0.50]\text{ fm}^{-3}$, $n_Q \in [0.50, 1.35]\text{ fm}^{-3}$ with $n_Q > n_H$ for Poly5; $\bar{n} \in [1.5, 8.0]\,n_0$, $\Gamma \in [2.0, 10.0]$ for boundary-free crossover), restricted to the set that yields a causal, thermodynamically stable EOS ($c_s^2\le 1$, $dP/dn>0$). The integral is estimated by Monte Carlo with $N_\psi=500$ draws. The Bayesian evidence $\mathcal{Z}$ and parameter posteriors for each transition scenario are computed using the \textit{dynesty} nested sampling package with 2000 live points and slice sampling~\cite{Skilling2006}. The relative statistical preference between competing transition paradigms is quantified by the log Bayes factor $\Delta\ln\mathcal{Z}_{X} = \ln\mathcal{Z}_X - \ln\mathcal{Z}_{base}$.

\section{Result}
\label{sec:result}

%%%%%%%%%%%%%%%%%%%%%%%%%%%%%%%%%%%%%%%%%%%%%%%%%%%%%%%
\begin{figure*}[t]\centering
\includegraphics[width=\textwidth]{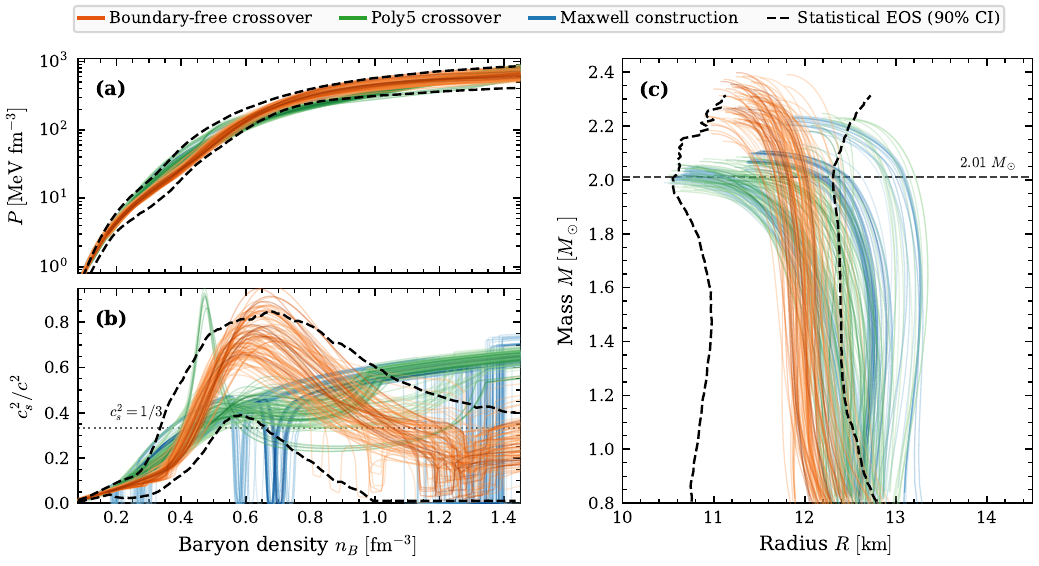}
\caption{Posterior EOSs and neutron-star properties for boundary-free crossover (orange), Poly5 crossover (green), and Maxwell construction (blue), compared against the statistical EOS 90\% credible interval (black dashed curves)~\cite{Huang:2026jgl}. Curves are reconstructed from posterior draws of $\bm{\phi}$ with their maximum a posteriori connection parameters $\bm{\psi}_{\rm MAP}$, with color saturation scaled logarithmically by posterior weight (darker curves denote higher probability density). (a) Pressure $P$ versus baryon density $n_B$. (b) Speed of sound squared $c_s^2$, with the conformal limit $c_s^2=1/3$ (horizontal dotted line). (c) Mass--radius relations, with the $2.01\,M_\odot$ lower bound from PSR J0740+6620~\cite{Fonseca:2021wxt} (horizontal dashed line).}
\label{fig:eos_mr}
\end{figure*}
%%%%%%%%%%%%%%%%%%%%%%%%%%%%%%%%%%%%%%%%%%%%%%%%%%%%%%%

Figure~\ref{fig:eos_mr} presents the representative EOS and mass--radius relations alongside the non-parametric statistical EOS posterior~\cite{Huang:2026jgl}. Representative curves are reconstructed from posterior draws of $\bm{\phi}$ by pairing each draw with its maximum a posteriori connection parameters $\bm{\psi}_{\rm MAP}$. Curves are colored by model family (orange for boundary-free, green for Poly5, blue for Maxwell), with color depth mapped logarithmically to posterior sample weight. Confronting microscopic candidate models directly with the statistical EOS posterior in correlated $(P, c_s^2)$ space reveals decisive evidence favoring the boundary-free crossover over the reference Maxwell construction ($\ln\mathcal{Z}_{\rm Maxwell} = -31.06 \pm 0.22$) by
\begin{equation}
\Delta\ln\mathcal{Z}_{\rm Boundary-free} = +7.20 \pm 0.28 \,,
\end{equation}
compared to $\Delta\ln\mathcal{Z}_{\rm Poly5} = +2.04 \pm 0.30$ for the Poly5 crossover. This also yields a decisive preference of $\Delta\ln\mathcal{Z} = +5.16 \pm 0.26$ for the boundary-free crossover over the polynomial interpolation, demonstrating that the intermediate stiffening in dense matter is governed by independent geometric connection parameters rather than the specific microscopic interactions of isolated hadronic and quark models.

The thermodynamic behavior associated with this preference is illustrated in the sound-speed profile (Fig.~\ref{fig:eos_mr}(b)). The statistical EOS posterior favors a profile characterized by a prominent peak ($c_s^2 > 1/3$) at intermediate densities, followed by extended softening with $c_s^2 < 1/3$, which has been well-reproduced by the boundary-free crossover. In contrast, the endpoint-matching conditions of the Poly5 crossover, together with
the thermodynamic identity
$\int_{\mu_H}^{\mu_Q} n\,d\mu = P_Q - P_H$,
constrain its sound-speed profile
within the specified polynomial ansatz.
Reproducing the intermediate sound-speed peak tends to favor a stiffer quark branch that conflicts with
the high-density softening ($c_s^2 < 1/3$).

Crucially, the incorporation of the bag constant $B$ amplifies the non-monotonic structure in the sound speed of the boundary-free crossover compared to the baseline without a bag constant ($B=0$) (See more discussions in Appendix C). In the energy-density interpolation, the extra pressure over the hadron-quark mixing takes the form $P_{\rm extra} = n_B w'(n_B)[\varepsilon_Q - \varepsilon_H]$. Because $B$ elevates the deconfined quark energy density ($\varepsilon_Q = \varepsilon_Q^{(0)} + B$), it contributes positively to $P_{\rm extra}$ throughout the crossover regime, and it enhances the sound-speed peak at intermediate densities ($n_B \sim 3$--$5\,n_0$). Simultaneously, the $-B$ offset in quark pressure ($P_Q = P_Q^{(0)} - B$) lowers the pressure at high densities ($w \to 1$, $w' \to 0$). The bag constant $B$ is therefore higher in the boundary-free crossover over the Maxwell and Poly5 construction, and supports a higher maximum mass of NS $M_{\rm TOV} = 2.27^{+0.22}_{-0.18} \ M_{\odot}$ than $\sim2.1 M_{\odot}$, which is more consistent with the result considering the NS mass function \cite{Fan:2023spm}.

As shown in Fig.~\ref{fig:eos_mr}(b), the PDM also has a first-order transition associated with the onset of $N(1535)$. A smaller chiral invariant mass $M_0$ shifts this transition to lower densities.
In the Maxwell construction, the neutron-star constraints favor smaller $M_0$, placing the hadronic transition at $n_B \sim 0.5$--$0.7\,\mathrm{fm}^{-3}$. When this transition occurs
before the hadron--quark transition, the EOS contains two first-order transitions, each with a constant-pressure interval where $c_s^2=0$. By contrast, the boundary-free crossover favors larger $M_0$, delaying the transition in the hadronic EOS to
$n_B \gtrsim 0.7\,\mathrm{fm}^{-3}$. The quark EOS has a larger weight in the interpolation at these densities. The sound speed of the combined EOS depends on both baseline EOSs and derivatives of
the interpolation weight, so it drops to a small but non-zero value. It hints that a realistic chiral restoration is not necessarily a phase transition. The posterior EOSs shown in Fig.~\ref{fig:eos_mr}(b) have positive $c_s^2$, with a peak of $c_s^2 \approx 0.7$ at $n_B \approx 3.8\,n_0$, followed by softening
to $c_s^2 \sim 0.3$--$0.4$. This behavior agrees more closely with the statistical EOS posterior. For Poly5, the transition in the hadronic EOS lies within
$[n_H,n_Q]$. The EOS in this interval is specified by the fifth-order polynomial rather than the hadronic model, so the onset of $N(1535)$
is not explicitly represented in the sound-speed profile.

%%%%%%%%%%%%%%%%%%%%%%%%%%%%%%%%%%%%%%%%%%%%%%%%%%%%%%%
\begin{figure}[t]\centering
\includegraphics[width=\columnwidth]{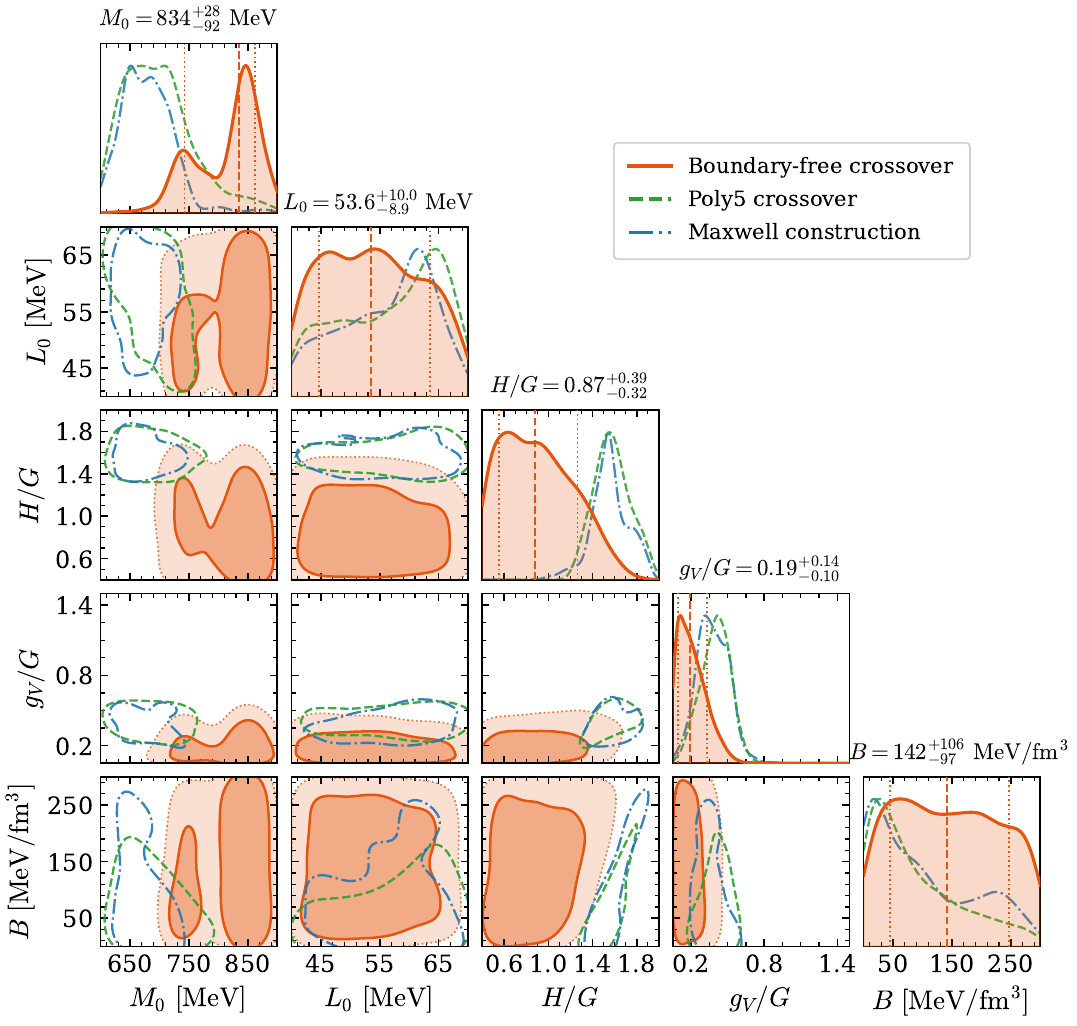}
\caption{Posterior distributions of microscopic EOS parameters ($M_0, L_0, H/G, g_V/G, B$) for boundary-free crossover (orange filled), Poly5 crossover (green dashed), and Maxwell construction (blue dot-dashed). Diagonal panels show 1D marginal posteriors with medians and $1\sigma$ credible intervals for boundary-free crossover. Off-diagonal panels show 2D joint credible regions (68\% and 95\% for boundary-free; 68\% for Poly5 crossover and Maxwell construction).}
\label{fig:params}
\end{figure}
%%%%%%%%%%%%%%%%%%%%%%%%%%%%%%%%%%%%%%%%%%%%%%%%%%%%%%%

These distinct transition mechanisms directly determine the inferred microscopic parameters of the hadron and quark phases (Fig.~\ref{fig:params}). The boundary-free crossover exhibits a pronounced preference for a large invariant mass of
\begin{equation}
M_0 = 834.4^{+28.2}_{-92.4}\text{ MeV} \,,
\end{equation}
compared to $M_0 = 677.5^{+41.0}_{-37.6}\text{ MeV}$ for Maxwell and $690.9^{+60.7}_{-51.6}\text{ MeV}$ for Poly5. 

A corresponding difference is shown in the quark sector for the diquark coupling $H/G$, which quantifies the strength of attractive pairing correlations responsible for color superconductivity. The Maxwell and Poly5 constructions favor relatively large diquark couplings, peaking at $H/G = 1.55^{+0.20}_{-0.14}$ and $1.57^{+0.18}_{-0.16}$, respectively. In Maxwell construction, a large $H/G$ is necessary for a valid crossing $P_H=P_Q$, since at low $H/G$ the quark pressure never matches the hadronic one and no first-order transition occurs. For Poly5, the thermodynamic integral constraint requires a large pressure jump across $[n_H, n_Q]$ to maintain stability, forcing the quark sector to be stiff. In contrast, the boundary-free crossover favors a smaller diquark coupling,
\begin{equation}
H/G = 0.87^{+0.39}_{-0.32} \,.
\end{equation}
The inferred bag constant is $B = 141.8^{+105.7}_{-97.2}\text{ MeV/fm}^3$ for the boundary-free crossover, compared with $74.8^{+139.2}_{-66.9}\text{ MeV/fm}^3$ for Maxwell construction and $62.2^{+112.0}_{-46.8}\text{ MeV/fm}^3$ for Poly5. The latter two constructions favor smaller values which lead to stiffer EOS for the quark matter.

In the three-flavor NJL-type quark model,  color superconductivity involves three pairing channels with gaps $\Delta_{ud} = -2 H d_s$, $\Delta_{us} = -2 H d_d$, and $\Delta_{ds} = -2 H d_u$. By solving the full 10-variable stationarity system under color and electric charge neutrality at highest densities of the constructed EOSs, the self-consistent solutions yield non-zero gaps that are nearly degenerate ($\Delta_{ud} \approx \Delta_{us} = \Delta_{ds}$ with a splitting below 5\%). This demonstrates that dense quark matter spontaneously realizes the CFL phase. We define the flavor-averaged CFL pairing gap as $\Delta_{\rm CFL} = (\Delta_{ud} + \Delta_{us} + \Delta_{ds})/3 = -2H(d_u + d_d + d_s)/3$. Early canonical reviews anticipated the color-superconducting gap at NS densities to lie in the range $\Delta \sim 10$--$100\text{ MeV}$~\cite{Alford:2007xm}, although subsequent phenomenological models tuning to massive pulsars often explored values exceeding $200\text{ MeV}$. 

\begin{figure}[t]\centering
\includegraphics[width=\linewidth]{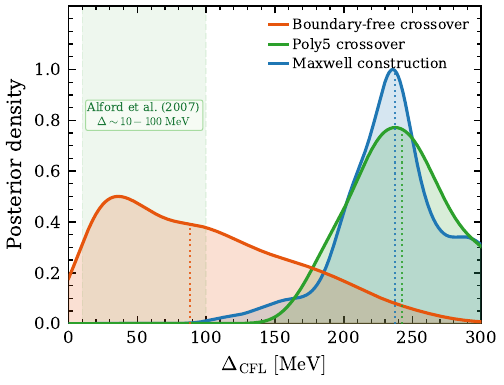}
\caption{Posterior probability distributions of the CFL pairing gap $\Delta_{\rm CFL}$ at $n_B \approx 8\,n_0$ for boundary-free crossover (orange), Poly5 crossover (green), and Maxwell construction (blue). Dashed vertical lines indicate medians; shaded green band denotes the theoretical expectation $\Delta \sim 10$--$100\text{ MeV}$~\cite{Alford:2007xm}.}
\label{fig:pairing_gap}
\end{figure}

The resulting posterior distributions $P(\Delta_{\rm CFL})$ are shown in Fig.~\ref{fig:pairing_gap}. For the Maxwell construction and Poly5 crossover, the pairing gap peaks at $\Delta_{\rm CFL} = 237.2^{+37.1}_{-32.5}\text{ MeV}$ and $242.2^{+32.1}_{-37.7}\text{ MeV}$, respectively. In contrast, the boundary-free crossover yields
\begin{equation}
\Delta_{\rm CFL} = 88.0^{+83.2}_{-58.5}\text{ MeV} \,.
\end{equation}
This result is consistent with the standard theoretical expectation $\Delta \sim 10$--$100\text{ MeV}$ summarized in the review by Alford et al.~\cite{Alford:2007xm}. Crucially, it is also in close agreement with recent constraints on the CFL pairing gap obtained from perturbative QCD matching at $\mu_B \approx 2.6\text{ GeV}$ ($\Delta^*_{\rm CFL} \lesssim 66\text{ MeV}$, with a posterior peak at $\sim 34\text{ MeV}$)~\cite{Tang:2026pqc}. In contrast, the excessively large gaps ($\Delta_{\rm CFL} \gtrsim 200\text{ MeV}$) demanded by the Maxwell and Poly5 frameworks are in strong tension with these limits.

%%%%%%%%%%%%%%%%%%%%%%%%%%%%%%%%%%%%%%%%%%%%%%%%%%%%%%%
%%%%%%%%%%%%%%%%%%%%%%%%%%%%%%%%%%%%%%%%%%%%%%%%%%%%%%%
\section{Summary}
\label{sec:summary}

In this Letter, using non-parametric statistical EOS posteriors derived from multi-messenger observations as data, we performed a Bayesian model comparison among hadron-quark transition paradigms connecting the PDM and NJL models. Confronting candidate equations of state directly in correlated $(P, c_s^2)$ space reveals decisive evidence favoring a continuous crossover over the conventional Maxwell construction, with the boundary-free formulation being strongly preferred. This demonstrates that dense matter in the transition region must deviate significantly from extrapolations of isolated effective models, where the intermediate stiffening originates from the geometric overlap related to hadron size rather than the interactions.

Beyond statistical preference, the favored boundary-free crossover exhibits intrinsic physical self-consistency. In the hadronic sector, it naturally accommodates a large chiral invariant nucleon mass ($M_0 \approx 834\text{ MeV}$) that delays chiral restoration to preserve hadronic dominance at early crossover. In the quark sector, it realizes a CFL phase with a small pairing gap ($\Delta_{\rm CFL} \approx 88\text{ MeV}$) consistent with perturbative QCD limits (considering small density-dependence from $\sim$1700 MeV in this work and 2600 MeV in the perturbative regime), avoiding the excessively large gaps ($\Delta \gtrsim 200\text{ MeV}$) demanded by conventional Maxwell or Poly5 constructions. The Bayesian evidence favoring the crossover over the conventional Maxwell construction does not rule out a genuine QCD phase boundary, which can still emerge from a spinodal instability within this unified framework. Nevertheless, in this picture, the hadron–quark transition in cold dense matter is fundamentally continuous, providing a self-consistent foundation to explore phase boundaries in dense matter rather than artificially connecting separate models.

%%%%%%%%%%%%%%%%%%%%%%%%%%%%%%%%%%%%%%
\begin{acknowledgments}
The analyses were carried out on the Hokusai Bigwaterfall supercomputer in RIKEN. Y.H is supported by the Postdoctoral Fellowship Program (No. GZC20241915) of the China Postdoctoral Science Foundation, the Project for Young Scientists in Basic Research (No. YSBR-088) of the Chinese Academy of Sciences, and the National Natural Science Foundation of China under Grants (No. 12588101 and No. 12233011). B.G. is supported in part by JSPS KAKENHI Grant No. 26K17147.
\end{acknowledgments}
%%%%%%%%%%%%%%%%%%%%%%%%%%%%%%%%%%%%%%%
\clearpage

\appendix
\section{EQUATION OF STATE}
\label{sec-eos}

In this section, we present the microscopic formulations of dense neutron star matter in the hadronic and deconfined quark phases, followed by the explicit construction of the unified equation of state (EOS) across the hadron-quark transition.

\subsection{Nuclear matter EOS: Parity Doublet Model}
\label{sec:PDM matter}

Throughout this work, we assume that hyperons do not emerge in the hadronic phase and restrict our analysis to two light flavors ($N_f = 2$). Following Ref.~\cite{Motohiro:2015taa}, the thermodynamic potential of the parity doublet model (PDM) is given by
\begin{equation}
\begin{aligned}
\Omega_{\mathrm{PDM}}= &\, V(\sigma) - V(\sigma_0) - \frac{1}{2} m_\omega^2 \omega^2 - \frac{1}{2} m_\rho^2 \rho^2 \\
&\, - \lambda_{\omega \rho}\left(g_\omega \omega\right)^2\left(g_\rho \rho\right)^2 \\
&\, - 2 \sum_{i=\pm} \sum_{\alpha=p, n} \int^{k_F^\alpha} \frac{\mathrm{d}^3 \mathbf{p}}{(2 \pi)^3}\left(\mu_\alpha^*-E_{\mathbf{p}}^i\right),
\end{aligned}
\end{equation}
where $i = \pm$ denotes the parity state of the nucleon ($N(939)$ for positive parity and $N(1535)$ for negative parity), and $E_{\mathbf{p}}^i = \sqrt{\mathbf{p}^2 + m_i^2}$ represents the single-particle baryon energy. The effective chemical potentials are $\mu_p^* = \mu_p - g_\omega \omega - \frac{1}{2} g_\rho \rho$ and $\mu_n^* = \mu_n - g_\omega \omega + \frac{1}{2} g_\rho \rho$. The chiral scalar potential $V(\sigma)$ takes the standard form
\begin{equation}
V(\sigma) = -\frac{1}{2}\bar{\mu}^2\sigma^2 + \frac{1}{4}\lambda_4 \sigma^4 -\frac{1}{6}\lambda_6\sigma^6 - m_\pi^2 f_\pi\sigma,
\end{equation}
where $\sigma_0 = f_\pi = 93\text{ MeV}$ is the vacuum expectation value breaking chiral symmetry spontaneously. Leptonic contributions from electrons and muons are incorporated via
\begin{equation}
\begin{aligned}
\Omega_{\rm H} =&\, \Omega_{\rm PDM} + \sum_{l = e, \mu}\Omega_l, \\
\Omega_l =&\, -2 \int^{k_F^l} \frac{\mathrm{d}^3 \mathbf{p}}{(2 \pi)^3}\left(\mu_l - \sqrt{\mathbf{p}^2 + m_l^2}\right),
\end{aligned}
\end{equation}
subject to $\beta$-equilibrium $\mu_n = \mu_p + \mu_e$, $\mu_\mu = \mu_e$, and electric charge neutrality $n_p = n_e + n_\mu$.

In the PDM, the nucleon $N(939)$ and its chiral partner $N(1535)$ acquire their masses through both the spontaneous chiral condensate $\sigma$ and an intrinsic chiral-invariant mass $M_0$:
\begin{equation}
m_\pm = \sqrt{M_0^2 + \left(\frac{g_1 + g_2}{2}\sigma\right)^2} \mp \frac{g_1 - g_2}{2} \sigma,
\label{eq:PDM-mass}
\end{equation}
where $g_1$ and $g_2$ are Yukawa couplings. In the chiral restoration limit ($\sigma \to 0$), the masses of the two parity partners become degenerate at $m_+ = m_- = M_0$.
The model parameters are calibrated to empirical nuclear matter properties at saturation density $n_0 = 0.16\text{ fm}^{-3}$: binding energy $E_0 / A = -16.0\text{ MeV}$, incompressibility $K_0 = 240\text{ MeV}$, symmetry energy $J_0 = 31.0\text{ MeV}$, and symmetry energy slope $L_0 \in [40, 70]\text{ MeV}$~\cite{Gao:2024chh, Motohiro:2015taa}. The slope $L_0$ is systematically adjusted by tuning the $\omega$-$\rho$ cross-coupling $\lambda_{\omega\rho}$ alongside the $\rho$-meson coupling $g_\rho$, controlling the density dependence of the symmetry energy without altering symmetric nuclear matter properties.

\begin{figure}[htbp]\centering
\includegraphics[width=0.9\hsize]{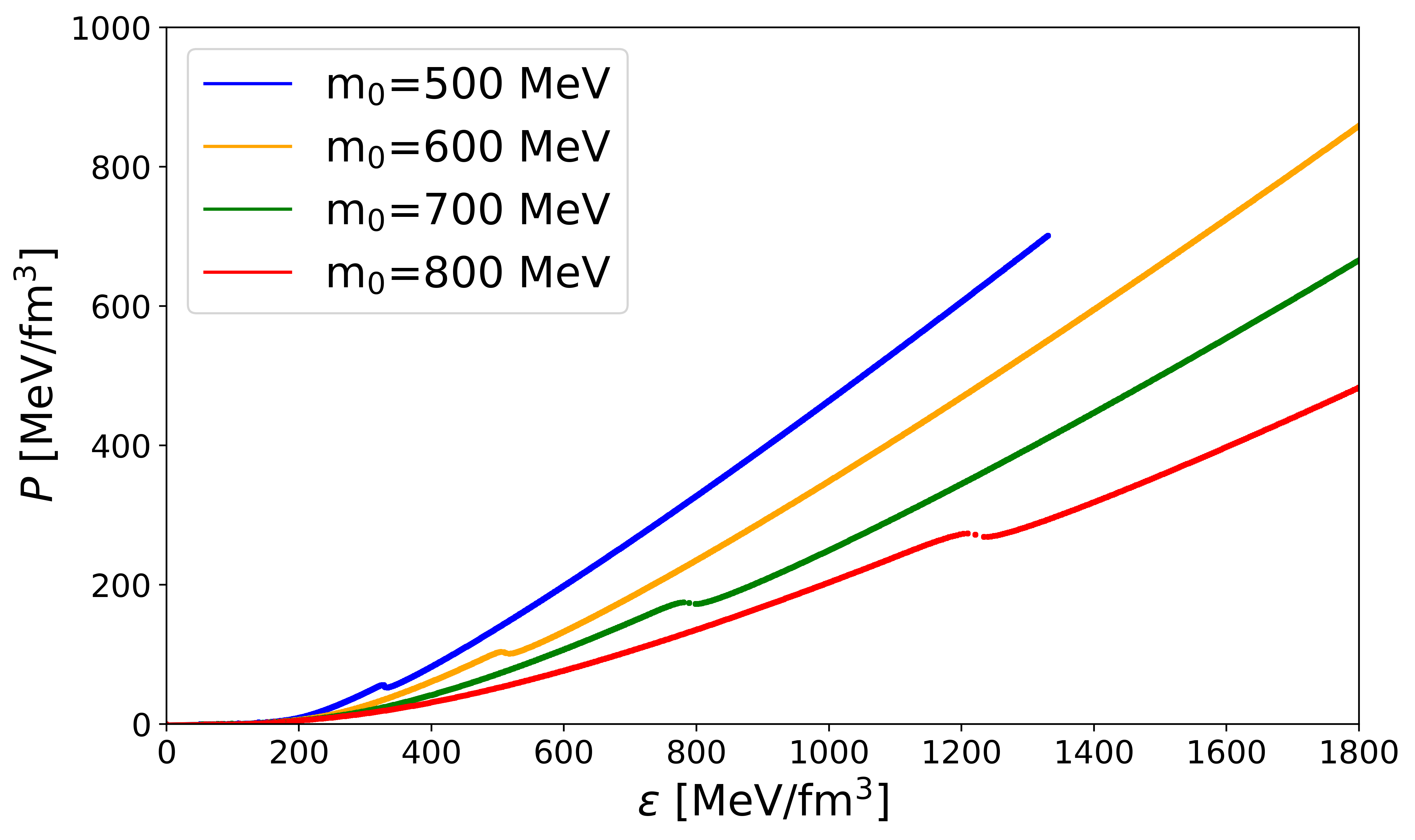}
\caption{Hadronic equations of state from the parity doublet model for representative values of the chiral invariant mass $M_0 \in [500, 800]\text{ MeV}$. The small kink at supranuclear densities reflects the threshold onset of the negative-parity nucleon partner $N(1535)$.}
\label{PDM_EOS}
\end{figure}

The chiral invariant mass $M_0$ acts as a crucial lever controlling the stiffness of hadronic matter. As $M_0$ increases, a smaller portion of the vacuum nucleon mass originates from the chiral condensate $\sigma_0$, which weakens the scalar Yukawa coupling $g_\sigma = (g_1 + g_2)/2$. Because reproducing the empirical binding energy at $n_0$ requires balancing scalar attraction against vector repulsion, the vector coupling $g_\omega$ also decreases monotonically with increasing $M_0$. At supranuclear densities where vector meson exchange dominates the EOS, a smaller $g_\omega$ results in substantially softer matter, as illustrated in Fig.~\ref{PDM_EOS}. Furthermore, the onset of the negative-parity partner $N(1535)$ at supranuclear density opens an additional Fermi sea, inducing a soft plateau in the hadronic EOS.

%%%%%%%%%%%%%%%%%%%%%%%%%%%%%%%%%%%%%%%%%%%%%%%%%%%%%%%
%%%%%%%%%%%%%%%%%%%%%%%%%%%%%%%%%%%%%%%%%%%%%%%%%%%%%%%

\subsection{Quark matter EOS: Three-flavor NJL model}
\label{NJL matter}

Following Refs.~\cite{Baym:2017whm, Baym:2019iky, Hatsuda:1994pi}, we describe deconfined quark matter using a three-flavor Nambu--Jona-Lasinio (NJL) model incorporating chiral symmetry breaking, vector repulsion, and color superconductivity:
\begin{equation}
\begin{aligned}
\mathcal{L}_{\rm NJL} =&\, \bar{q}(i\gamma^\mu \partial_\mu - \hat{m})q \\
&\, + G \sum_{a=0}^8 \left[(\bar{q}\lambda_a q)^2 + (\bar{q}i\gamma_5\lambda_a q)^2\right] \\
&\, - K \left[ \det\nolimits_f \bar{q}(1+\gamma_5)q + \det\nolimits_f \bar{q}(1-\gamma_5)q \right] \\
&\, + H \sum_{A, A'=2,5,7} \left[(\bar{q} i\gamma_5 \tau_A \lambda_{A'} q_c)(\bar{q}_c i\gamma_5 \tau_A \lambda_{A'} q)\right] \\
&\, - g_V (\bar{q}\gamma^\mu q)^2,
\end{aligned}
\end{equation}
where $\hat{m} = \mathrm{diag}(m_u, m_d, m_s)$ are current quark masses, $\lambda_a$ and $\tau_A$ denote Gell-Mann matrices in flavor and color spaces, and $q_c = C\bar{q}^T$ is the charge-conjugate spinor. The parameters are fixed to standard Hatsuda-Kunihiro values reproducing vacuum meson phenomenology~\cite{Hatsuda:1994pi}: $\Lambda = 631.4\text{ MeV}$, $G\Lambda^2 = 1.835$, $K\Lambda^5 = 9.29$, $m_u = m_d = 5.5\text{ MeV}$, and $m_s = 135.7\text{ MeV}$. The vector coupling $g_V$ and diquark coupling $H$ are treated as free parameters governing the stiffness and pairing in dense quark matter.

In the mean-field approximation, the thermodynamic potential of dense quark matter with color superconductivity takes the form
\begin{equation}
\Omega_{\mathrm{CSC}} = \Omega_s - \Omega_s^{\rm vac} + \Omega_c - \Omega_c^{\rm vac},
\end{equation}
where
\begin{align}
\Omega_s &= - \sum_{i=1}^{18} \int^\Lambda \frac{\mathrm{d}^3 \mathbf{p}}{(2 \pi)^3} \epsilon_i(\mathbf{p}), \\
\Omega_c &= \sum_{i=u,d,s} (2 G \sigma_i^2 + H d_i^2) - 4 K \sigma_u \sigma_d \sigma_s - g_V n_q^2.
\end{align}
Here $\sigma_i = \langle\bar{q}_i q_i\rangle$ denote the chiral condensates, $d_i$ denote the diquark condensates, and $n_q = \langle q^\dagger q\rangle$ is the quark number density. The constituent quark masses and pairing gaps are given by
\begin{align}
M_i &= m_i - 4 G \sigma_i + 2 K |\epsilon_{ijk}| \sigma_j \sigma_k, \label{eq:M_uds}\\
\Delta_i &= -2 H d_i, \label{eq:gap_diquark}
\end{align}
where $(i, j, k)$ run over cyclic permutations of $(u, d, s)$. The single-particle quasiparticle excitation spectra $\epsilon_i(\mathbf{p})$ ($i = 1, \dots, 18$ distinct positive energy eigenvalues) are obtained by diagonalizing the inverse quark propagator in the $72 \times 72$ Nambu-Gorkov color-flavor-spin space:
\begin{equation}
S^{-1}(k) = \begin{pmatrix}
\gamma_\mu k^\mu - \hat{M} + \gamma^0 \hat{\mu} & \gamma_5 \sum_{i=1}^3 \Delta_i R_i \\
-\gamma_5 \sum_{i=1}^3 \Delta_i^* R_i & \gamma_\mu k^\mu - \hat{M} - \gamma^0 \hat{\mu}
\end{pmatrix},
\end{equation}
with $(R_1, R_2, R_3) = (\tau_7\lambda_7, \tau_5\lambda_5, \tau_2\lambda_2)$, $\hat{M} = \mathrm{diag}(M_u, M_d, M_s)$, and the effective chemical potential matrix:
\begin{equation}
\hat{\mu} = (\mu_q - 2 g_V n_q) \mathbb{I} + \mu_3 \lambda_3 + \mu_8 \lambda_8 + \mu_Q Q,
\end{equation}
where $Q = \mathrm{diag}(2/3, -1/3, -1/3)$ is the electric charge generator, and $\mu_3, \mu_8$ are color chemical potentials.

Cold quark matter in neutron stars is subject to weak-interaction $\beta$-equilibrium, electric charge neutrality, and color charge neutrality. The equilibrium state is uniquely determined by solving the 10-dimensional stationarity system for the state vector
\begin{equation}
\mathbf{X} = (\sigma_u, \sigma_d, \sigma_s, d_u, d_d, d_s, n_q, \mu_3, \mu_8, \mu_Q)^T.
\end{equation}
The 10 equations comprise:
\begin{enumerate}
\item Three chiral gap equations:
\begin{equation}
\frac{\partial \Omega_{\rm Q}}{\partial \sigma_i} = 0 \quad (i = u, d, s),
\end{equation}
\item Three diquark gap equations:
\begin{equation}
\frac{\partial \Omega_{\rm Q}}{\partial d_i} = 0 \quad (i = u, d, s),
\end{equation}
yielding pairing gaps $\Delta_i = -2 H d_i$. At high densities, the system relaxes into the CFL phase where color and flavor symmetries are locked into a diagonal subgroup, resulting in nearly degenerate pairing gaps $\Delta_u \approx \Delta_d \approx \Delta_s \equiv \Delta_{\rm CFL} = -\frac{2}{3} H (d_u + d_d + d_s)$.
\item The vector mean-field consistency relation:
\begin{equation}
\frac{\partial \Omega_{\rm Q}}{\partial n_q} = 0 \implies n_q = - \frac{\partial \Omega_s}{\partial \mu_q^*},
\end{equation}
where $\mu_q^* = \mu_q - 2 g_V n_q$ represents the shifted quark chemical potential.
\item Two color neutrality constraints:
\begin{equation}
n_3 = - \frac{\partial \Omega_{\rm Q}}{\partial \mu_3} = 0, \quad n_8 = - \frac{\partial \Omega_{\rm Q}}{\partial \mu_8} = 0,
\end{equation}
guaranteeing vanishing net color isospin and color hypercharge.
\item Electromagnetic charge neutrality under $\beta$-equilibrium:
\begin{equation}
n_Q^{\rm tot} = - \frac{\partial \Omega_{\rm Q}}{\partial \mu_Q} = n_Q^{\rm quark} - n_e - n_\mu = 0,
\end{equation}
where the lepton chemical potentials satisfy $\mu_e = \mu_\mu = -\mu_Q$, and the lepton number densities are $n_l = \frac{1}{3\pi^2}(\mu_l^2 - m_l^2)^{3/2}\Theta(\mu_l - m_l)$ for $l = e, \mu$.
\end{enumerate}
Once the system is solved at each $\mu_q = \mu_B / 3$, the base pressure and energy density are obtained via $P_{\rm Q}^{(0)} = -\Omega_{\rm Q}$ and $\varepsilon_{\rm Q}^{(0)} = -P_{\rm Q}^{(0)} + \mu_B n_B + \sum_l \mu_l n_l$ with $n_B = n_q / 3$. Incorporating the phenomenological bag constant $B$, the physical quark pressure and energy density entering the unified equations of state are shifted according to $P_{\rm Q} = P_{\rm Q}^{(0)} - B$ and $\varepsilon_{\rm Q} = \varepsilon_{\rm Q}^{(0)} + B$.

\section{Hadron-quark transition paradigms}
\label{sec:QH-transition}

With the hadronic EOS constructed from the PDM (Sec.~\ref{sec:PDM matter}) and the quark EOS from the NJL model (Sec.~\ref{NJL matter}), we combine them into unified equations of state covering the entire density range realized in neutron stars. In the low-density crust regime $n_B \lesssim 0.5\, n_0$, all models are seamlessly matched to the BPS (Baym-Pethick-Sutherland) crust EOS~\cite{Baym:1971pw}. At supranuclear densities, we examine three representative transition paradigms: (i) a first-order phase transition via Maxwell construction, (ii) a fixed-window polynomial crossover, and (iii) a boundary-free energy-density crossover described by smooth transition weighting.

\subsection{First-order phase transition: Maxwell construction}
\label{sec:Maxwell}

When the hadron-quark transition is first order between locally charge-neutral phases, the transition is implemented via the Maxwell construction. Phase coexistence is established at the critical baryon chemical potential $\mu_B^{\rm c}$ where the grand potentials (and hence pressures) of the two phases intersect:
\begin{equation}
P_{\rm H}(\mu_B^{\rm c}) = P_{\rm Q}(\mu_B^{\rm c}).
\label{eq:Maxwell_condition}
\end{equation}
Thermodynamic stability dictates that the ground state selects the phase of maximal pressure at any given chemical potential:
\begin{equation}
P(\mu_B) = \max\left[ P_{\rm H}(\mu_B), P_{\rm Q}(\mu_B) \right].
\label{eq:Maxwell_max}
\end{equation}
At $\mu_B^{\rm c}$, the baryon density exhibits a finite discontinuity $\Delta n_B = n_B^{\rm Q} - n_B^{\rm H}$ with $n_B = \mathrm{d}P/\mathrm{d}\mu_B$, accompanied by a latent heat discontinuity in energy density $\Delta\varepsilon = \mu_B^{\rm c}\Delta n_B$. Across this two-phase density gap $[n_B^{\rm H}, n_B^{\rm Q}]$, pressure remains strictly constant ($dP/d\varepsilon = 0$), causing the sound speed to vanish identically ($c_s^2 = 0$).

\subsection{Poly5 crossover: Polynomial interpolation}
\label{sec:poly}

If the transition proceeds through a continuous crossover within an intermediate window, the two descriptions can be bridged by a polynomial interpolation~\cite{Baym:2017whm, Baym:2019iky}. The segmentation of the polynomial crossover EOS is summarized in Table~\ref{UniEOS}.

\begin{table}[tbh]
\begin{center}
\begin{tabular}{c|c|c|c}
\hline
\hline
$0\leq \hat{n}<0.5$ & $0.5\leq \hat{n}\leq \hat{n}_H$ & $\hat{n}_H <\hat{n}< \hat{n}_Q$ & $\hat{n}\geq \hat{n}_Q$\\
\hline
\rm{Crust} & \rm{PDM} & \rm{Interpolation} & \rm{NJL}\\
\hline
\hline
\end{tabular}
\end{center}
\caption{Segmentation and composition of the polynomial crossover EOS, with the normalized density $\hat{n}=n_B/n_0$ and the normalized matching densities $\hat{n}_{H,Q}=n_{H,Q}/n_0$.}
\label{UniEOS}
\end{table}

The pressure in the crossover window $[\mu_H, \mu_Q]$ is expressed as a fifth-order polynomial:
\begin{equation}
P_{\mathrm{I}}(\mu_{B}) = \sum_{k=0}^{5} C_{k} \mu_{B}^{k},
\label{eq:polyP}
\end{equation}
where the six coefficients $C_k$ are uniquely fixed by demanding $C^2$ matching at the window endpoints:
\begin{equation}
\begin{aligned}
\left.\frac{\mathrm{d}^{n} P_{\mathrm{I}}}{\mathrm{d} \mu_{B}^{n}}\right|_{\mu_{H}} &= \left.\frac{\mathrm{d}^{n} P_{\mathrm{H}}}{\mathrm{d} \mu_{B}^{n}}\right|_{\mu_{H}}, \\
\left.\frac{\mathrm{d}^{n} P_{\mathrm{I}}}{\mathrm{d} \mu_{B}^{n}}\right|_{\mu_{Q}} &= \left.\frac{\mathrm{d}^{n} P_{\mathrm{Q}}}{\mathrm{d} \mu_{B}^{n}}\right|_{\mu_{Q}}, \quad (n = 0, 1, 2),
\end{aligned}
\label{eq:polyBC}
\end{equation}
matching pressure, baryon density $n_B = \mathrm{d}P/\mathrm{d}\mu_B$, and baryon susceptibility $\chi_B = \mathrm{d}^2P/\mathrm{d}\mu_B^2$ at the matching boundaries $\mu_H$ and $\mu_Q$, corresponding to the transition densities $n_H$ and $n_Q$, respectively.
The interpolated EOS must satisfy thermodynamic stability ($\chi_B \ge 0$) and causality ($c_s^2 = n_B / (\mu_B \chi_B) \le 1$).

To establish why polynomial interpolations frequently suffer from mechanical instabilities, we present the exact mathematical structure of the interpolated density profile across $[\mu_H, \mu_Q]$.

\vspace{0.5em}
\noindent\textbf{Theorem (Quartic Density Decomposition Identity).} Let $x = (\mu_B - \mu_H)/\Delta\mu \in [0, 1]$ be the normalized chemical potential coordinate, where $\Delta\mu = \mu_Q - \mu_H > 0$. The baryon density $n_{\rm I}(\mu_B) = \mathrm{d}P_{\rm I}/\mathrm{d}\mu_B$ is a quartic polynomial in $x$ that admits the unique, exact decomposition
\begin{equation}
n_{\rm I}(x) = n_{\rm cubic}(x) + \Delta\langle n\rangle\, b(x),
\qquad
b(x)\equiv 30\,x^{2}(1-x)^{2},
\label{eq:poly5_decomposition}
\end{equation}
where
\begin{equation}
\Delta\langle n\rangle \equiv \int_0^1\!\big[n_{\rm I}(x)-n_{\rm cubic}(x)\big]\,\mathrm{d}x
= \langle n_{\rm I}\rangle - \langle n_{\rm cubic}\rangle
\end{equation}
is the integral density mismatch, and $n_{\rm cubic}(x)$ is the unique cubic Hermite interpolant matching the four boundary values and derivatives:
\begin{equation}
\begin{aligned}
n_{\rm cubic}(x) =&\, n_H (1 - 3x^2 + 2x^3) + n_Q (3x^2 - 2x^3) \\
&\, + \Delta\mu\,\chi_H x (1 - x)^2 - \Delta\mu\,\chi_Q x^2 (1 - x),
\end{aligned}
\label{eq:hermite_cubic}
\end{equation}
with $\chi_H = \left.\mathrm{d}n/\mathrm{d}\mu_B\right|_{\mu_H}$ and $\chi_Q = \left.\mathrm{d}n/\mathrm{d}\mu_B\right|_{\mu_Q}$. The integral mean densities of the cubic background and of the full interpolant are, respectively,
\begin{equation}
\langle n_{\rm cubic}\rangle \equiv \int_0^1 n_{\rm cubic}(x)\,\mathrm{d}x = \frac{n_H + n_Q}{2} + \frac{\Delta\mu}{12}\left(\chi_H - \chi_Q\right)
\label{eq:nbar_cubic}
\end{equation}
and
\begin{equation}
\langle n_{\rm I}\rangle \equiv \int_0^1 n_{\rm I}(x)\,\mathrm{d}x = \frac{P_{\rm Q} - P_{\rm H}}{\mu_Q - \mu_H} = \frac{\Delta P}{\Delta\mu},
\label{eq:nbar_thermo}
\end{equation}
the latter following directly from the pressure-matching conditions.

\vspace{0.5em}
\noindent\textbf{Proof.} Because $P_{\rm I}(\mu_B)$ is a fifth-order polynomial, its derivative $n_{\rm I}(x)$ is a polynomial of degree at most four. At the boundaries $x = 0$ and $x = 1$, $n_{\rm I}(x)$ satisfies $n_{\rm I}(0) = n_H$, $n_{\rm I}'(0) = \Delta\mu\,\chi_H$, $n_{\rm I}(1) = n_Q$, and $n_{\rm I}'(1) = \Delta\mu\,\chi_Q$, which are identically matched by $n_{\rm cubic}(x)$. Consequently, the difference $\delta(x) \equiv n_{\rm I}(x) - n_{\rm cubic}(x)$ satisfies $\delta(0) = \delta'(0) = \delta(1) = \delta'(1) = 0$. Since $\delta(x) \in \mathcal{P}_4$ possesses double roots at both endpoints, it must take the form $\delta(x) = c\, x^2 (1 - x)^2$ for some scalar constant $c$. Integrating over $x \in [0, 1]$ and using the Euler Beta integral $\int_0^1 x^2 (1 - x)^2\,\mathrm{d}x = \mathrm{B}(3, 3) = \frac{\Gamma(3)\Gamma(3)}{\Gamma(6)} = \frac{1}{30}$, we obtain
\begin{equation}
\int_0^1 \delta(x)\,\mathrm{d}x = \frac{c}{30} = \langle n_{\rm I}\rangle - \langle n_{\rm cubic}\rangle = \Delta\langle n\rangle,
\end{equation}
which fixes $c = 30\,\Delta\langle n\rangle$ uniquely. $\blacksquare$

\vspace{0.5em}
The identity (\ref{eq:poly5_decomposition}) reveals the mathematical origin of the spinodal instability in Poly5 constructions and explains why the generation of a sound-speed peak is fundamentally coupled to the quark EOS. Differentiating Eq.~(\ref{eq:poly5_decomposition}) with respect to $\mu_B$, the baryon number susceptibility within the crossover window is
\begin{equation}
\chi_B(x) = \frac{1}{\Delta\mu}\left[ n_{\rm cubic}'(x) + \Delta\langle n\rangle\, b'(x) \right],
\label{eq:poly5_chi}
\end{equation}
where $b'(x)\equiv \mathrm{d}b/\mathrm{d}x$; the same bubble polynomial $b(x)$ appears in Eq.~(\ref{eq:poly5_decomposition}). The squared speed of sound is then given by
\begin{equation}
c_s^2(x) = \frac{n_{\rm I}(x)}{\mu_B(x)\,\chi_B(x)} = \frac{\Delta\mu\,n_{\rm I}(x)}{\mu_B(x)\left[ n_{\rm cubic}'(x) + \Delta\langle n\rangle\, b'(x) \right]}.
\label{eq:poly5_cs2}
\end{equation}
Crucially, the derivative of the bubble polynomial,
\begin{equation}
b'(x) = 60\,x(1-x)(1-2x),
\end{equation}
is anti-symmetric about the midpoint $x = 0.5$. It attains an extremum of opposite sign in each half of the interpolation window: a positive maximum $b'_{\rm max} = 10/\sqrt{3} \approx 5.77$ at $x_1 = (3-\sqrt{3})/6 \approx 0.211$, and a negative minimum $b'_{\rm min} = -5.77$ at $x_2 = (3+\sqrt{3})/6 \approx 0.789$. This sign reversal explains why $c_s^2$ generation in Poly5 is intrinsically hostage to the quark EOS.

First, thermodynamic stability imposes a strict constraint on the entrance region ($x < 0.5$). Because $b'(x) > 0$ in this interval, a negative mismatch $\Delta\langle n\rangle < 0$ suppresses the susceptibility. To maintain thermodynamic stability with $\chi_B \ge 0$, the total slope $n_{\rm I}'(x)$ must remain non-negative throughout $x \in [0, 1]$. This requirement establishes a strict lower bound on the mismatch:
\begin{equation}
\Delta\langle n\rangle \ge - \min_{x \in (0, 0.5)} \left[ \frac{n_{\rm cubic}'(x)}{60\, x(1-x)(1-2x)} \right] \approx -0.08\text{ fm}^{-3}.
\label{eq:deltan_crit}
\end{equation}

Second, the mismatch $\Delta\langle n\rangle$ is inherently coupled to the quark EOS. In the Poly5 formulation, $\Delta\langle n\rangle$ is not an independent internal parameter, but is fixed by the thermodynamic integral condition $\Delta\langle n\rangle = (P_{\rm Q} - P_{\rm H})/\Delta\mu - \langle n_{\rm cubic}\rangle$. In physical quark models with moderate diquark pairing ($H/G \sim 0.87$, corresponding to realistic pairing gaps $\Delta_{\rm CFL} \lesssim 90\text{ MeV}$ consistent with perturbative QCD), the quark pressure $P_{\rm Q}$ is moderate. This yields $\Delta P \approx 77\text{ MeV/fm}^3$ and a large negative mismatch $\Delta\langle n\rangle \approx -0.35\text{ fm}^{-3}$.

Third, this coupling renders the sound-speed peak hostage to the quark EOS. To avoid spinodal instability and simultaneously suppress $\chi_B(x)$ in the upper window to generate a stiff sound-speed peak ($c_s^2 > 1/3$) capable of supporting $2\,M_\odot$ neutron stars, the Poly5 scheme must artificially inflate the quark boundary pressure $P_{\rm Q}$ to reach $\Delta P \gtrsim 160\text{ MeV/fm}^3$. This requirement strictly demands exceedingly large diquark couplings ($H/G \sim 1.50$--$1.60$, $\Delta_{\rm CFL} \sim 240\text{ MeV}$).

Consequently, within the Poly5 scheme, the emergence of a sound-speed peak cannot occur as an independent physical property of hadron percolation; rather, it is directly dictated by and coupled to the high-density quark EOS. Artificially freezing the quark sector into an overly stiff branch ($c_s^2 \sim 0.5$) to secure stability during interpolation prevents dense matter from softening towards the conformal limit ($c_s^2 \to 1/3$) at higher densities.

\subsection{Boundary-free crossover: Weight function interpolation}
\label{sec:boundary_free}

In contrast to polynomial matching in chemical potential space, the boundary-free crossover formulation directly interpolates the energy density of the two phases in density space $n_B$~\cite{Masuda:2012kf, Masuda:2012ed}:
\begin{equation}
\varepsilon_{\mathrm{I}}(n_B) = \left[1 - w(n_B)\right] \varepsilon_{\mathrm{H}}(n_B) + w(n_B) \varepsilon_{\mathrm{Q}}(n_B),
\label{eq:masuda_E}
\end{equation}
where $\varepsilon_{\rm H}(n_B)$ and $\varepsilon_{\rm Q}(n_B)$ are the energy densities of charge-neutral, $\beta$-equilibrated hadronic matter (from the PDM) and quark matter (from the NJL model), respectively. The function $w(n_B)$ is a smooth, monotonic weight function with connection parameters $(\bar{n}, \Gamma)$ controlling the crossover center density and sharpness.

Thermodynamic consistency dictates that the chemical potential is obtained from the fundamental relation $\mu_B = \mathrm{d}\varepsilon_{\rm I}/\mathrm{d}n_B$:
\begin{equation}
\begin{aligned}
\mu_B(n_B) =&\, \left[1 - w(n_B)\right] \mu_B^{\rm H}(n_B) + w(n_B) \mu_B^{\rm Q}(n_B) \\
&\, + w'(n_B) \left[\varepsilon_{\mathrm{Q}}(n_B) - \varepsilon_{\mathrm{H}}(n_B)\right],
\end{aligned}
\label{eq:masuda_mu}
\end{equation}
where $\mu_B^{\rm H} = \mathrm{d}\varepsilon_{\rm H}/\mathrm{d}n_B$ and $\mu_B^{\rm Q} = \mathrm{d}\varepsilon_{\rm Q}/\mathrm{d}n_B$. The pressure is then determined by the zero-temperature Euler relation $P_{\rm I} = \mu_B n_B - \varepsilon_{\rm I}$:
\begin{equation}
\begin{aligned}
P_{\mathrm{I}}(n_B) =&\, \left[1 - w(n_B)\right] P_{\mathrm{H}}(n_B) + w(n_B) P_{\mathrm{Q}}(n_B) \\
&\, + n_B w'(n_B) \left[\varepsilon_{\mathrm{Q}}(n_B) - \varepsilon_{\mathrm{H}}(n_B)\right].
\end{aligned}
\label{eq:masuda_P}
\end{equation}
Crucially, thermodynamic consistency automatically generates the emergent term
\begin{equation}
P_{\rm extra}(n_B) \equiv n_B w'(n_B) \left[\varepsilon_{\mathrm{Q}}(n_B) - \varepsilon_{\mathrm{H}}(n_B)\right].
\label{eq:masuda_Pextra}
\end{equation}
Because the crossover weight is monotonically increasing ($w' > 0$) and deconfined quark matter has a higher energy density than hadronic matter at the same baryon density ($\varepsilon_{\rm Q} > \varepsilon_{\rm H}$), this emergent contribution is strictly positive ($P_{\rm extra} > 0$). It elevates the blended pressure above the baseline envelope of both isolated models, naturally generating a prominent sound-speed peak ($c_s^2 > 1/3$) at intermediate densities ($n_B \sim 3$--$4\,n_0$). This stiffening reflects the physical repulsion induced by nucleon geometric overlap and quark Pauli blocking during percolation, rather than artificial interactions.

Different from earlier phenomenological crossover models that assumed symmetric switching forms (such as hyperbolic-tangent or symmetric logistic functions~\cite{Masuda:2012kf, Masuda:2012ed}), we adopt an asymmetric exponential weight function. This choice is physically grounded in the profound scale hierarchy of dense QCD matter: while the density interval from nuclear saturation $n_0 \approx 0.16\text{ fm}^{-3}$ to the central core of maximum-mass neutron stars spans merely $\sim 5\,n_0$, the subsequent evolution from neutron star cores to the asymptotic domain of perturbative QCD spans $> 20\,n_0$. We therefore parameterize the crossover using an asymmetric exponential weight:
\begin{equation}
w(n_B) = \exp\left[ -\left(\frac{\bar{n}}{n_B}\right)^\Gamma \right],
\label{eq:exp_weight}
\end{equation}
with derivative
\begin{equation}
w'(n_B) = \frac{\Gamma}{\bar{n}} \left(\frac{\bar{n}}{n_B}\right)^{\Gamma + 1} \exp\left[ -\left(\frac{\bar{n}}{n_B}\right)^\Gamma \right].
\end{equation}
This asymmetric weight possesses essential physical and mathematical virtues: as $n_B \to 0$, $w(n_B) \to 0$ exponentially fast, ensuring that nuclear matter around saturation density $n_0$ remains purely hadronic with negligible quark admixture ($w(n_0) \sim 10^{-7}$); as $n_B \to \infty$, $w(n_B) \to 1$ smoothly, recovering pure quark matter asymptotically without artificial reflection onto the broad perturbative QCD scale; and its derivative maintains a smooth, single-peaked structure without the $(1-2w)$ zero-crossing dip typical of symmetric logistic forms.

Because this energy-density crossover formulation requires no polynomial matching conditions at arbitrary boundaries, it is completely free from the spinodal instabilities that plague the Poly5 scheme. The intermediate stiffening is controlled entirely by the geometric percolation scale $\Gamma$, decoupling the crossover peak from the stiffness of the soft quark phase. Consequently, the quark sector is free to adopt physical diquark couplings ($H/G \approx 0.87$, $\Delta_{\rm CFL} \approx 88\text{ MeV}$), enabling dense matter to undergo extended softening deeply below the conformal value ($c_s^2 < 1/3$).

The three constructions introduced above share the same microscopic inputs, using the PDM at low density and the NJL model at high density. However, they embody qualitatively different physical paradigms for the emergence of quark degrees of freedom. The conventional Maxwell construction implements an abrupt two-phase separation characterized by a finite density gap. The polynomial interpolation describes a smooth crossover confined within a fixed matching window. Finally, the energy-density interpolation formulates a unified crossover without sharp phase boundaries.

Each construction defines a distinct family of hybrid or crossover EOSs governed by the shared microscopic parameters $(M_0, L_0, H/G, g_V/G, B)$, while adopting fundamentally different parameterizations of the transition region. For the Maxwell construction, the transition chemical potential $\mu_B^{\rm c}$ is dynamically fixed by the thermodynamic phase equilibrium $P_{\rm H}(\mu_B^{\rm c}) = P_{\rm Q}(\mu_B^{\rm c})$, introducing no additional transition parameters. For the polynomial crossover (Poly5), the interpolation is bounded by the free matching densities $(n_H, n_Q)$ with $n_Q > n_H$, sampled over $n_H \in [0.24, 0.50]\text{ fm}^{-3}$ and $n_Q \in [0.50, 1.35]\text{ fm}^{-3}$. For the boundary-free crossover, the unified EOS is governed by the independent geometric connection parameters $(\bar{n}, \Gamma)$, sampled over $\bar{n} \in [1.5, 8.0]\,n_0$ and $\Gamma \in [2.0, 10.0]$. In our Bayesian model comparison, these model-specific transition parameters $\bm{\psi} = (n_H, n_Q)$ and $\bm{\psi} = (\bar{n}, \Gamma)$ are marginalized out as nuisance parameters, allowing an unbiased comparison among all three transition paradigms within the common microscopic parameter space $\bm{\phi} = (M_0, L_0, H/G, g_V/G, B)$.

\section{Baseline analysis with $B=0$}
\label{sec:B0_baseline}
To demonstrate that our model selection and microscopic inferences do not artifactually depend on the bag constant, we examine the baseline four-dimensional parameter space $\bm{\phi} = (M_0, L_0, H/G, g_V/G)$ with $B \equiv 0$. In this baseline configuration, the Bayesian evidence remains decisively in favor of the boundary-free crossover, yielding $\Delta\ln\mathcal{Z} = +6.20 \pm 0.27$ relative to the Maxwell construction and $+3.93 \pm 0.24$ relative to the Poly5 interpolation.

Setting $B=0$ removes the vacuum confinement energy shift, depriving the quark sector of both the positive energy-density offset $+B$ and the negative vacuum pressure shift $-B$. Under constraints, the inferred parameters in the boundary-free crossover shift slightly to $M_0 = 827.8^{+34.1}_{-98.7}\text{ MeV}$, $H/G = 0.76^{+0.30}_{-0.25}$, and $\Delta_{\rm CFL} = 67.5^{+61.6}_{-43.7}\text{ MeV}$ (compared to $M_0 = 686.4\text{ MeV}$, $H/G = 1.45$, $\Delta_{\rm CFL} = 217.4\text{ MeV}$ for Maxwell and $M_0 = 705.7\text{ MeV}$, $H/G = 1.44$, $\Delta_{\rm CFL} = 214.0\text{ MeV}$ for Poly5). This confirms that the qualitative preference for a large chiral invariant mass and physical diquark pairing is robust against the inclusion of $B$.

\begin{figure*}[htbp]
\centering
\includegraphics[width=\textwidth]{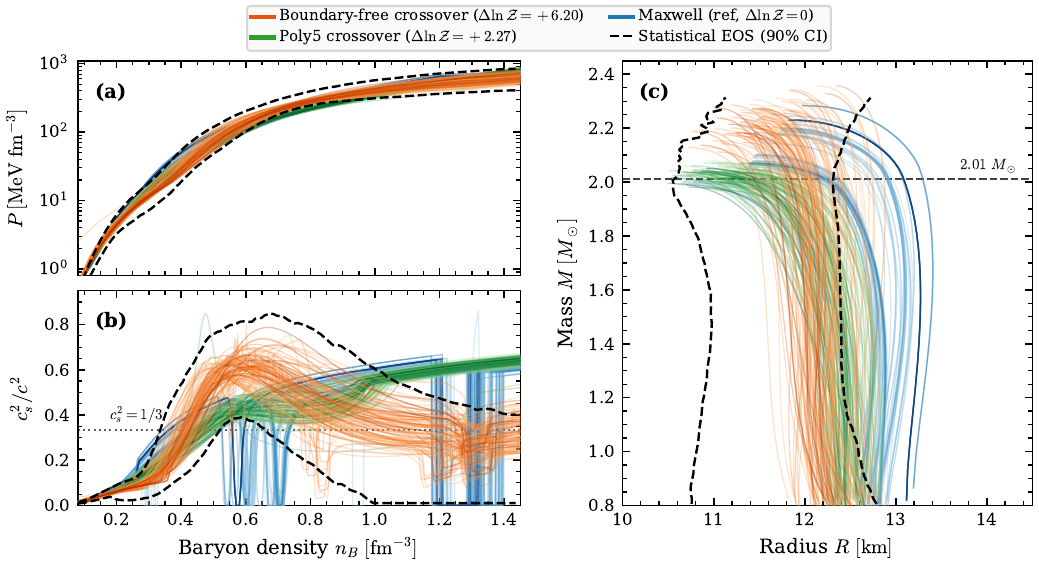} % 
\caption{Posterior equations of state and macroscopic neutron-star properties for the baseline analysis with vanishing bag constant ($B=0$). (a) Pressure $P$ versus baryon density $n_B$. (b) Speed of sound squared $c_s^2$, displaying a narrower and lower peak compared to the $B \ne 0$ case. (c) Mass--radius relations confronted with the $2.01\,M_\odot$ constraint from PSR J0740+6620. }
\label{fig:eos_mr_B0}
\end{figure*}

\begin{figure}[htbp]
\centering
\includegraphics[width=\columnwidth]{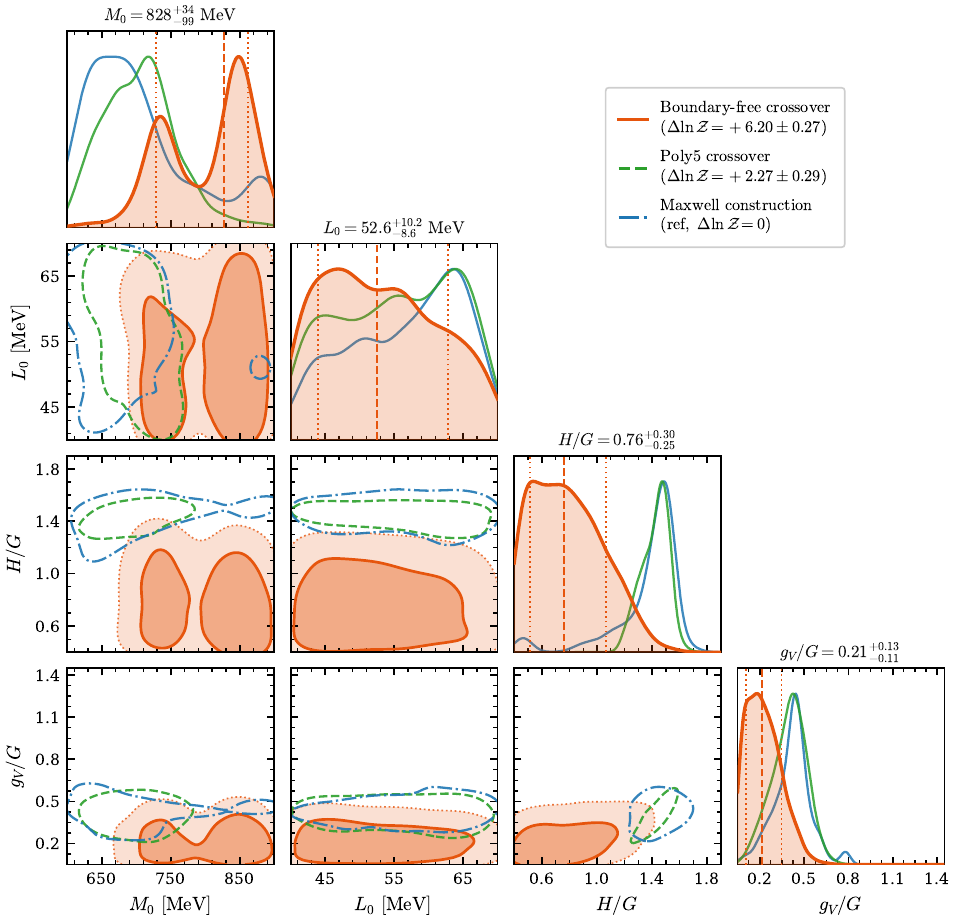} 
\caption{Posterior distributions of the four microscopic EOS parameters ($M_0, L_0, H/G, g_V/G$) for the baseline analysis ($B=0$), comparing boundary-free crossover (orange), Poly5 crossover (green), and Maxwell construction (blue). Diagonal panels display 1D marginal posteriors with medians and $1\sigma$ credible intervals for the boundary-free crossover; off-diagonal panels display 2D joint credible regions. Model-comparison Bayes factors $\Delta\ln\mathcal{Z}$ are indicated in the legend.}
\label{fig:params_B0}
\end{figure}

Crucially, introducing a finite bag constant $B > 0$ actively {\it amplifies} the non-monotonic structure of the speed of sound $c_s^2$. In the boundary-free crossover formulation, the interpolated energy density $\varepsilon_{\rm I} = (1-w)\varepsilon_{\rm H} + w\varepsilon_{\rm Q}$ yields the pressure
\begin{equation}
\begin{split}
P_{\rm I}(n_B) &= [1-w(n_B)] P_{\rm H}(n_B) + w(n_B) P_{\rm Q}(n_B) \\
&\quad + n_B w'(n_B) [\varepsilon_{\rm Q}(n_B) - \varepsilon_{\rm H}(n_B)].
\end{split}
\label{eq:masuda_P_app}
\end{equation}
With addtional contribution from the bag constant $\varepsilon_{\rm Q}(n_B) = \varepsilon_{\rm Q}^{(0)}(n_B) + B$ and $P_{\rm Q}(n_B) = P_{\rm Q}^{(0)}(n_B) - B$, Eq.~\ref{eq:masuda_P_app} separates into
\begin{equation}
P_{\rm I}(n_B) = P_{\rm I}(n_B) |_{B=0} + B \left[ n_B w'(n_B) - w(n_B) \right].
\label{eq:P_I_B_mod}
\end{equation}
Differentiating Eq.~\ref{eq:P_I_B_mod} with respect to baryon density, the cancellation of first derivatives yields the exact relation
\begin{equation}
\frac{\mathrm{d}P_{\rm I}}{\mathrm{d}n_B} = \left(\frac{\mathrm{d}P_{\rm I}}{\mathrm{d}n_B}\right)_{B=0} + B \, n_B \, w''(n_B).
\label{eq:dP_dn_B_mod}
\end{equation}
The squared speed of sound $c_s^2 = \frac{1}{\mu_B}\frac{\mathrm{d}P_{\rm I}}{\mathrm{d}n_B}$ is therefore modulated by the second derivative $w''(n_B)$:
\begin{equation}
c_s^2(n_B) = \frac{1}{\mu_B(n_B)}\left[ \left(\frac{\mathrm{d}P_{\rm I}}{\mathrm{d}n_B}\right)_{B=0} + B \, n_B \, w''(n_B) \right].
\label{eq:cs2_B_mod}
\end{equation}

The sign of this modulation is governed by the inflection point of the weight function $w(n_B) = \exp[-(\bar{n}/n_B)^\Gamma]$, given by $n_{\rm infl} \equiv \bar{n}[\Gamma/(\Gamma+1)]^{1/\Gamma}$ where $w''(n_{\rm infl}) = 0$. In the early percolation ascent ($n_B < n_{\rm infl}$), the accelerating weight ($w'' > 0$) provides a positive boost $B n_B w'' > 0$ that steepens $\mathrm{d}P/\mathrm{d}n_B$, heightening and sharpening the sound-speed peak ($c_s^2 > 1/3$) required to support $2\,M_\odot$ neutron stars. Conversely, beyond the inflection density ($n_B > n_{\rm infl}$), the decelerating weight ($w'' < 0$) turns $B n_B w'' < 0$ into a strong negative contribution that sharply depresses $\mathrm{d}P/\mathrm{d}n_B$, forcing $c_s^2$ down well below the conformal value ($c_s^2 < 1/3$) and driving dense core matter toward a spinodal turning point ($\mathrm{d}P/\mathrm{d}n_B \to 0$).

In the absence of the bag constant ($B=0$), this density-dependent curvature correction drops out, yielding a noticeably flatter sound-speed profile across the transition. A finite bag constant $B$ is therefore essential to actively amplify both the intermediate percolation stiffening and the high-density softening toward the spinodal threshold, providing a closer match to the non-parametric statistical EOS and increasing the Bayesian preference for the boundary-free crossover from $\Delta\ln\mathcal{Z} = +6.20$ to $+7.20$.

\bibliography{ref.bib}

@article{Yang:1952be,
    author = "Yang, C. N. and Lee, T. D.",
    title = "{Statistical Theory of Equations of State and Phase Transitions. I. Theory of Condensation}",
    doi = "10.1103/PhysRev.87.404",
    journal = "Phys. Rev.",
    volume = "87",
    pages = "404--410",
    year = "1952"
}

@article{Lee:1952pf,
    author = "Lee, T. D. and Yang, C. N.",
    title = "{Statistical Theory of Equations of State and Phase Transitions. II. Lattice Gas and Ising Model}",
    doi = "10.1103/PhysRev.87.410",
    journal = "Phys. Rev.",
    volume = "87",
    pages = "410--419",
    year = "1952"
}

@article{Lebowitz:1966,
    author = "Lebowitz, J. L. and Penrose, O.",
    title = "{Rigorous Treatment of the Van Der Waals-Maxwell Theory of the Liquid-Vapor Transition}",
    doi = "10.1063/1.1704821",
    journal = "J. Math. Phys.",
    volume = "7",
    pages = "98--113",
    year = "1966"
}

@article{Chomaz:2003dz,
    author = "Chomaz, Ph. and Colonna, M. and Randrup, J.",
    title = "{Nuclear spinodal fragmentation}",
    eprint = "nucl-th/0306030",
    archivePrefix = "arXiv",
    doi = "10.1016/j.physrep.2003.09.006",
    journal = "Phys. Rept.",
    volume = "389",
    pages = "263--440",
    year = "2004"
}

@article{Randrup:2009gp,
    author = "Randrup, Jorgen",
    title = "{Phase-transition dynamics for conserved charges}",
    eprint = "0903.4136",
    archivePrefix = "arXiv",
    primaryClass = "nucl-th",
    doi = "10.1103/PhysRevC.79.044901",
    journal = "Phys. Rev. C",
    volume = "79",
    pages = "044901",
    year = "2009"
}

@article{Baym:1979,
    author = "Baym, Gordon",
    title = "{Confinement of quarks in nuclear matter}",
    doi = "10.1016/0378-4371(79)90204-7",
    journal = "Physica A",
    volume = "96",
    pages = "131--135",
    year = "1979"
}

@article{Asakawa:1995zu,
    author = "Asakawa, M. and Hatsuda, T.",
    title = "{What thermodynamics tells about QCD plasma near phase transition}",
    eprint = "hep-ph/9508360",
    archivePrefix = "arXiv",
    doi = "10.1103/PhysRevD.55.4488",
    journal = "Phys. Rev. D",
    volume = "55",
    pages = "4488--4491",
    year = "1997"
}

@article{Hatsuda:1994pi,
    author = "Hatsuda, Tetsuo and Kunihiro, Teiji",
    title = "{QCD phenomenology based on a chiral effective Lagrangian}",
    eprint = "hep-ph/9401310",
    archivePrefix = "arXiv",
    reportNumber = "UTHEP-270, RYUTHP-94-1",
    doi = "10.1016/0370-1573(94)90022-1",
    journal = "Phys. Rept.",
    volume = "247",
    pages = "221--367",
    year = "1994"
}

@article{Baym:2017whm,
    author = "Baym, Gordon and Hatsuda, Tetsuo and Kojo, Toru and Powell, Philip D. and Song, Yifan and Takatsuka, Tatsuyuki",
    title = "{From hadrons to quarks in neutron stars: a review}",
    eprint = "1707.04966",
    archivePrefix = "arXiv",
    primaryClass = "astro-ph.HE",
    reportNumber = "RIKEN-ITHEMS-REPORT-17, RIKEN-QHP-316, RIKEN-iTHEMS-Report-17",
    doi = "10.1088/1361-6633/aaae14",
    journal = "Rept. Prog. Phys.",
    volume = "81",
    number = "5",
    pages = "056902",
    year = "2018"
}

@article{Detar:1988kn,
    author = "Detar, Carleton E. and Kunihiro, Teiji",
    title = "{Linear $\sigma$ Model With Parity Doubling}",
    reportNumber = "RIFP-748",
    doi = "10.1103/PhysRevD.39.2805",
    journal = "Phys. Rev. D",
    volume = "39",
    pages = "2805",
    year = "1989"
}

@article{Aarts:2015mma,
    author = {Aarts, Gert and Allton, Chris and Hands, Simon and J\"ager, Benjamin and Praki, Chrisanthi and Skullerud, Jon-Ivar},
    title = "{Nucleons and parity doubling across the deconfinement transition}",
    eprint = "1502.03603",
    archivePrefix = "arXiv",
    primaryClass = "hep-lat",
    doi = "10.1103/PhysRevD.92.014503",
    journal = "Phys. Rev. D",
    volume = "92",
    number = "1",
    pages = "014503",
    year = "2015"
}

@article{Aarts:2017rrl,
    author = {Aarts, Gert and Allton, Chris and De Boni, Davide and Hands, Simon and J\"ager, Benjamin and Praki, Chrisanthi and Skullerud, Jon-Ivar},
    title = "{Light baryons below and above the deconfinement transition: medium effects and parity doubling}",
    eprint = "1703.09246",
    archivePrefix = "arXiv",
    primaryClass = "hep-lat",
    doi = "10.1007/JHEP06(2017)034",
    journal = "JHEP",
    volume = "06",
    pages = "034",
    year = "2017"
}

@article{Aarts:2018glk,
    author = {Aarts, Gert and Allton, Chris and De Boni, Davide and J\"ager, Benjamin},
    title = "{Hyperons in thermal QCD: A lattice view}",
    eprint = "1812.07393",
    archivePrefix = "arXiv",
    primaryClass = "hep-lat",
    reportNumber = "CP3-Origins-2018-050 DNRF90",
    doi = "10.1103/PhysRevD.99.074503",
    journal = "Phys. Rev. D",
    volume = "99",
    number = "7",
    pages = "074503",
    year = "2019"
}

@article{Jido:2001nt,
    author = "Jido, Daisuke and Oka, Makoto and Hosaka, Atsushi",
    title = "{Chiral symmetry of baryons}",
    eprint = "hep-ph/0110005",
    archivePrefix = "arXiv",
    doi = "10.1143/PTP.106.873",
    journal = "Prog. Theor. Phys.",
    volume = "106",
    pages = "873--908",
    year = "2001"
}

@article{Motohiro:2015taa,
    author = "Motohiro, Yuichi and Kim, Youngman and Harada, Masayasu",
    title = "{Asymmetric nuclear matter in a parity doublet model with hidden local symmetry}",
    eprint = "1505.00988",
    archivePrefix = "arXiv",
    primaryClass = "nucl-th",
    doi = "10.1103/PhysRevC.92.025201",
    journal = "Phys. Rev. C",
    volume = "92",
    number = "2",
    pages = "025201",
    year = "2015",
    note = "[Erratum: Phys.Rev.C 95, 059903 (2017)]"
}

@article{Marczenko:2020jma,
    author = "Marczenko, Micha\l{} and Blaschke, David and Redlich, Krzysztof and Sasaki, Chihiro",
    title = "{Toward a unified equation of state for multi-messenger astronomy}",
    eprint = "2004.09566",
    archivePrefix = "arXiv",
    primaryClass = "astro-ph.HE",
    reportNumber = "CERN-TH-2020-061",
    doi = "10.1051/0004-6361/202038211",
    journal = "Astron. Astrophys.",
    volume = "643",
    pages = "A82",
    year = "2020"
}

@article{LIGOScientific:2017ync,
    author = "Abbott, B. P. and others",
    collaboration = "LIGO Scientific, Virgo, Fermi GBM, INTEGRAL, IceCube, AstroSat Cadmium Zinc Telluride Imager Team, IPN, Insight-Hxmt, ANTARES, Swift, AGILE Team, 1M2H Team, Dark Energy Camera GW-EM, DES, DLT40, GRAWITA, Fermi-LAT, ATCA, ASKAP, Las Cumbres Observatory Group, OzGrav, DWF (Deeper Wider Faster Program), AST3, CAASTRO, VINROUGE, MASTER, J-GEM, GROWTH, JAGWAR, CaltechNRAO, TTU-NRAO, NuSTAR, Pan-STARRS, MAXI Team, TZAC Consortium, KU, Nordic Optical Telescope, ePESSTO, GROND, Texas Tech University, SALT Group, TOROS, BOOTES, MWA, CALET, IKI-GW Follow-up, H.E.S.S., LOFAR, LWA, HAWC, Pierre Auger, ALMA, Euro VLBI Team, Pi of Sky, Chandra Team at McGill University, DFN, ATLAS Telescopes, High Time Resolution Universe Survey, RIMAS, RATIR, SKA South Africa/MeerKAT",
    title = "{Multi-messenger Observations of a Binary Neutron Star Merger}",
    eprint = "1710.05833",
    archivePrefix = "arXiv",
    primaryClass = "astro-ph.HE",
    reportNumber = "LIGO-P1700294, VIR-0802A-17, FERMILAB-PUB-17-478-A-AE-CD",
    doi = "10.3847/2041-8213/aa91c9",
    journal = "Astrophys. J. Lett.",
    volume = "848",
    number = "2",
    pages = "L12",
    year = "2017"
}

@article{LIGOScientific:2018cki,
    author = "Abbott, B. P. and others",
    collaboration = "LIGO Scientific, Virgo",
    title = "{GW170817: Measurements of neutron star radii and equation of state}",
    eprint = "1805.11581",
    archivePrefix = "arXiv",
    primaryClass = "gr-qc",
    reportNumber = "LIGO-P1800115",
    doi = "10.1103/PhysRevLett.121.161101",
    journal = "Phys. Rev. Lett.",
    volume = "121",
    number = "16",
    pages = "161101",
    year = "2018"
}

@article{Baym:2019iky,
    author = "Baym, Gordon and Furusawa, Shun and Hatsuda, Tetsuo and Kojo, Toru and Togashi, Hajime",
    title = "{New Neutron Star Equation of State with Quark-Hadron Crossover}",
    eprint = "1903.08963",
    archivePrefix = "arXiv",
    primaryClass = "astro-ph.HE",
    reportNumber = "RIKEN-iTHEMS-Report-19",
    doi = "10.3847/1538-4357/ab441e",
    journal = "Astrophys. J.",
    volume = "885",
    pages = "42",
    year = "2019"
}

@article{Kojo:2021wax,
    author = "Kojo, Toru and Baym, Gordon and Hatsuda, Tetsuo",
    title = "{Implications of NICER for Neutron Star Matter: The QHC21 Equation of State}",
    eprint = "2111.11919",
    archivePrefix = "arXiv",
    primaryClass = "astro-ph.HE",
    reportNumber = "RIKEN-iTHEMS-Report-21",
    doi = "10.3847/1538-4357/ac7876",
    journal = "Astrophys. J.",
    volume = "934",
    number = "1",
    pages = "46",
    year = "2022"
}

@article{Kojo:2020krb,
    author = "Kojo, Toru",
    title = "{QCD equations of state and speed of sound in neutron stars}",
    eprint = "2011.10940",
    archivePrefix = "arXiv",
    primaryClass = "nucl-th",
    doi = "10.1007/s43673-021-00011-6",
    journal = "AAPPS Bull.",
    volume = "31",
    number = "1",
    pages = "11",
    year = "2021"
}

@article{Kojo:2015fua,
    author = "Kojo, Toru",
    title = "{Phenomenological neutron star equations of state: 3-window modeling of QCD matter}",
    eprint = "1508.04408",
    archivePrefix = "arXiv",
    primaryClass = "hep-ph",
    doi = "10.1140/epja/i2016-16051-0",
    journal = "Eur. Phys. J. A",
    volume = "52",
    number = "3",
    pages = "51",
    year = "2016"
}

@article{Masuda:2012kf,
    author = "Masuda, Kota and Hatsuda, Tetsuo and Takatsuka, Tatsuyuki",
    title = "{Hadron-Quark Crossover and Massive Hybrid Stars with Strangeness}",
    eprint = "1205.3621",
    archivePrefix = "arXiv",
    primaryClass = "nucl-th",
    doi = "10.1088/0004-637X/764/1/12",
    journal = "Astrophys. J.",
    volume = "764",
    pages = "12",
    year = "2013"
}

@article{Masuda:2012ed,
    author = "Masuda, Kota and Hatsuda, Tetsuo and Takatsuka, Tatsuyuki",
    title = "{Hadron\textendash{}quark crossover and massive hybrid stars}",
    eprint = "1212.6803",
    archivePrefix = "arXiv",
    primaryClass = "nucl-th",
    doi = "10.1093/ptep/ptt045",
    journal = "PTEP",
    volume = "2013",
    number = "7",
    pages = "073D01",
    year = "2013"
}

@article{Baym:1971pw,
    author = "Baym, Gordon and Pethick, Christopher and Sutherland, Peter",
    title = "{The Ground state of matter at high densities: Equation of state and stellar models}",
    doi = "10.1086/151216",
    journal = "Astrophys. J.",
    volume = "170",
    pages = "299--317",
    year = "1971"
}

@article{Drischler:2020hwi,
    author = "Drischler, C. and Furnstahl, R. J. and Melendez, J. A. and Phillips, D. R.",
    title = "{How Well Do We Know the Neutron-Matter Equation of State at the Densities Inside Neutron Stars? A Bayesian Approach with Correlated Uncertainties}",
    eprint = "2004.07232",
    archivePrefix = "arXiv",
    primaryClass = "nucl-th",
    doi = "10.1103/PhysRevLett.125.202702",
    journal = "Phys. Rev. Lett.",
    volume = "125",
    number = "20",
    pages = "202702",
    year = "2020"
}

@article{McLerran:2018hbz,
    author = "McLerran, Larry and Reddy, Sanjay",
    title = "{Quarkyonic Matter and Neutron Stars}",
    eprint = "1811.12503",
    archivePrefix = "arXiv",
    primaryClass = "nucl-th",
    reportNumber = "INT-PUB-18-060",
    doi = "10.1103/PhysRevLett.122.122701",
    journal = "Phys. Rev. Lett.",
    volume = "122",
    number = "12",
    pages = "122701",
    year = "2019"
}

@article{Gao:2024chh,
    author = "Gao, Bikai and Yan, Yan and Harada, Masayasu",
    title = "{Reconciling constraints from the supernova remnant HESS J1731-347 with the parity doublet model}",
    eprint = "2404.04786",
    archivePrefix = "arXiv",
    primaryClass = "nucl-th",
    doi = "10.1103/PhysRevC.109.065807",
    journal = "Phys. Rev. C",
    volume = "109",
    number = "6",
    pages = "065807",
    year = "2024"
}

@article{Fukushima:2013rx,
    author = "Fukushima, Kenji and Sasaki, Chihiro",
    title = "{The phase diagram of nuclear and quark matter at high baryon density}",
    eprint = "1301.6377",
    archivePrefix = "arXiv",
    primaryClass = "hep-ph",
    doi = "10.1016/j.ppnp.2013.05.003",
    journal = "Prog. Part. Nucl. Phys.",
    volume = "72",
    pages = "99--154",
    year = "2013"
}

@article{Huang:2026jgl,
    author = "Huang, Yong-Jia and Tang, Shao-Peng and Fan, Yi-Zhong",
    title = "{The Non-parametric Equation of State Realizes a Generalized Quark-Hadron Crossover}",
    journal = "arXiv e-prints",
    eprint = "2605.08584",
    archivePrefix = "arXiv",
    primaryClass = "astro-ph.HE",
    reportNumber = "RIKEN-iTHEMS-Report-26",
    month = "5",
    year = "2026"
}

@article{Tang:2026pqc,
    author = "Tang, Shao-Peng and Huang, Yong-Jia and Fan, Yi-Zhong",
    title = "{Constraining the Color-Superconducting Pairing Gap with Perturbative QCD and Neutron Star Observations}",
    journal = "arXiv preprint",
    eprint = "2606.03707",
    year = "2026"
}

@article{Bazavov:2014pvz,
    author = "Bazavov, A. and others",
    collaboration = "HotQCD",
    title = "{Equation of state in ( 2+1 )-flavor QCD}",
    eprint = "1407.6387",
    archivePrefix = "arXiv",
    primaryClass = "hep-lat",
    reportNumber = "BNL-105928-2014-JA",
    doi = "10.1103/PhysRevD.90.094503",
    journal = "Phys. Rev. D",
    volume = "90",
    pages = "094503",
    year = "2014"
}

@article{Fonseca:2021wxt,
    author = "Fonseca, E. and others",
    title = "{Refined Mass and Geometric Measurements of the High-Mass PSR J0740+6620}",
    eprint = "2104.00880",
    archivePrefix = "arXiv",
    primaryClass = "astro-ph.HE",
    doi = "10.3847/2041-8213/ac03b8",
    journal = "Astrophys. J. Lett.",
    volume = "915",
    number = "1",
    pages = "L12",
    year = "2021"
}

@article{Riley:2019yda,
    author = "Riley, Thomas E. and others",
    title = "{A NICER View of PSR J0030+0451: Millisecond Pulsar Parameter Estimation}",
    eprint = "1912.05702",
    archivePrefix = "arXiv",
    primaryClass = "astro-ph.HE",
    doi = "10.3847/2041-8213/ab481c",
    journal = "Astrophys. J. Lett.",
    volume = "887",
    number = "1",
    pages = "L21",
    year = "2019"
}

@article{Miller:2019cac,
    author = "Miller, M. C. and others",
    title = "{PSR J0030+0451 Mass and Radius from $NICER$ Data and Implications for the Properties of Neutron Star Matter}",
    eprint = "1912.05705",
    archivePrefix = "arXiv",
    primaryClass = "astro-ph.HE",
    doi = "10.3847/2041-8213/ab50c5",
    journal = "Astrophys. J. Lett.",
    volume = "887",
    number = "1",
    pages = "L24",
    year = "2019"
}

@article{Komoltsev:2021jzg,
    author = "Komoltsev, Oleg and Kurkela, Aleksi",
    title = "{How Perturbative QCD Constrains the Equation of State at Neutron-Star Densities}",
    eprint = "2111.05350",
    archivePrefix = "arXiv",
    primaryClass = "nucl-th",
    doi = "10.1103/PhysRevLett.128.202701",
    journal = "Phys. Rev. Lett.",
    volume = "128",
    number = "20",
    pages = "202701",
    year = "2022"
}

@article{Alford:2007xm,
    author = "Alford, Mark G. and Schmitt, Andreas and Rajagopal, Krishna and Sch{\"a}fer, Thomas",
    title = "{Color superconductivity in dense quark matter}",
    eprint = "0709.4635",
    archivePrefix = "arXiv",
    primaryClass = "hep-ph",
    doi = "10.1103/RevModPhys.80.1455",
    journal = "Rev. Mod. Phys.",
    volume = "80",
    pages = "1455--1515",
    year = "2008"
}

@article{Oka:1980ax,
    author = {Oka, M. and Yazaki, K.},
    title = {{Nuclear Force in a Quark Model}},
    journal = {Phys. Lett. B},
    volume = {90},
    pages = {41--44},
    year = {1980},
    doi = {10.1016/0370-2693(80)90046-5}
}

@article{Dexheimer:2009hi,
    author = {Dexheimer, V. A. and Schramm, S.},
    title = {{A Novel Approach to Model Hybrid Stars}},
    eprint = {0901.1748},
    archivePrefix = {arXiv},
    primaryClass = {astro-ph.SR},
    doi = {10.1103/PhysRevC.81.045201},
    journal = {Phys. Rev. C},
    volume = {81},
    pages = {045201},
    year = {2010}
}

@article{Horowitz:1985gv,
    author = {Horowitz, C. J. and Moniz, E. J. and Negele, J. W.},
    title = {{Hadron structure in a simple model of quark/nuclear matter}},
    journal = {Phys. Rev. D},
    volume = {31},
    pages = {1689--1699},
    year = {1985},
    doi = {10.1103/PhysRevD.31.1689}
}

@article{Ropke:1986qs,
    author = {Ropke, G. and Blaschke, D. and Schulz, H.},
    title = {{Pauli Quenching Effects in a Simple String Model of Quark / Nuclear Matter}},
    journal = {Phys. Rev. D},
    volume = {34},
    pages = {3499--3513},
    year = {1986},
    doi = {10.1103/PhysRevD.34.3499}
}

@article{Lee:2023ofg,
    author = "Lee, Su Houng",
    title = "{Chiral Symmetry Breaking and the Masses of Hadrons: A Review}",
    eprint = "2303.14415",
    archivePrefix = "arXiv",
    primaryClass = "hep-ph",
    doi = "10.3390/sym15040799",
    journal = "Symmetry",
    volume = "15",
    number = "4",
    pages = "799",
    year = "2023"
}

@article{Kim:2020zae,
    author = "Kim, Jisu and Lee, Su Houng",
    title = "{Vector meson mass in the chiral symmetry restored vacuum}",
    eprint = "2012.06463",
    archivePrefix = "arXiv",
    primaryClass = "nucl-th",
    doi = "10.1103/PhysRevD.103.L051501",
    journal = "Phys. Rev. D",
    volume = "103",
    number = "5",
    pages = "L051501",
    year = "2021"
}

@article{Ji:1994av,
    author = "Ji, Xiang-Dong",
    title = "{A QCD analysis of the mass structure of the nucleon}",
    eprint = "hep-ph/9410274",
    archivePrefix = "arXiv",
    reportNumber = "MIT-CTP-2368",
    doi = "10.1103/PhysRevLett.74.1071",
    journal = "Phys. Rev. Lett.",
    volume = "74",
    pages = "1071--1074",
    year = "1995"
}

@article{Gao:2025nkg,
    author = "Gao, Bikai and Liu, Xiang and Harada, Masayasu and Ma, Yong-Liang",
    title = "{Implication of neutron star observations to the origin of nucleon mass}",
    eprint = "2508.00243",
    archivePrefix = "arXiv",
    primaryClass = "nucl-th",
    doi = "10.1007/s11433-025-2839-7",
    journal = "Sci. China Phys. Mech. Astron.",
    volume = "69",
    number = "3",
    pages = "232011",
    year = "2026"
}

@article{Gao:2026scv,
    author = "Gao, Bikai",
    title = "{Chiral symmetry restoration and hyperon suppression in neutron stars}",
    eprint = "2602.12503",
    archivePrefix = "arXiv",
    primaryClass = "nucl-th",
    doi = "10.1103/vb5r-vdm7",
    journal = "Phys. Rev. D",
    volume = "113",
    number = "8",
    pages = "083012",
    year = "2026"
}

@article{Kunihiro:2026gjo,
    author = "Kunihiro, Teiji",
    title = "{Chiral Symmetry and Its Restoration in QCD}",
    journal = "arXiv e-prints",
    eprint = "2604.27982",
    archivePrefix = "arXiv",
    primaryClass = "nucl-th",
    reportNumber = "YITP-26-49",
    month = "4",
    year = "2026"
}

@article{Geng:2026hbf,
    author = "Geng, Xuesong and Huang, Kaixuan and Shen, Hong and Li, Lei and Hu, Jinniu",
    title = "{Crossover equation of state constrained by astronomical observations and pQCD}",
    eprint = "2604.08841",
    archivePrefix = "arXiv",
    primaryClass = "nucl-th",
    doi = "10.1103/lb59-lwkr",
    journal = "Phys. Rev. D",
    volume = "113",
    number = "10",
    pages = "103002",
    year = "2026"
}

@article{Kawaguchi:2025cuf,
    author = "Kawaguchi, Mamiya and Harada, Masayasu and Ma, Yong-Liang",
    title = "{Origin of hadron mass from gravitational D-form factor and neutron star measurements}",
    eprint = "2512.23937",
    archivePrefix = "arXiv",
    primaryClass = "hep-ph",
    doi = "10.1016/j.physletb.2026.140400",
    journal = "Phys. Lett. B",
    volume = "876",
    pages = "140400",
    year = "2026"
}

@article{Xia:2024wpz,
    author = "Xia, Cheng-Jun",
    title = "{Extended NJL model for baryonic matter and quark matter}",
    eprint = "2405.02946",
    archivePrefix = "arXiv",
    primaryClass = "nucl-th",
    doi = "10.1103/PhysRevD.110.014022",
    journal = "Phys. Rev. D",
    volume = "110",
    number = "1",
    pages = "014022",
    year = "2024"
}

@article{Li:2019ztm,
    author = "Li, Cheng-Ming and Zuo, Shu-Yu and Yan, Yan and Zhao, Ya-Peng and Wang, Fei and Huang, Yong-Feng and Zong, Hong-Shi",
    title = "{Strange quark stars within proper time regularized (2+1)-flavor NJL model}",
    eprint = "1912.05093",
    archivePrefix = "arXiv",
    primaryClass = "hep-ph",
    doi = "10.1103/PhysRevD.101.063023",
    journal = "Phys. Rev. D",
    volume = "101",
    number = "6",
    pages = "063023",
    year = "2020"
}

@article{Salmi:2024aum,
    author = "Salmi, Tuomo and others",
    title = "{The Radius of the High-mass Pulsar PSR J0740+6620 with 3.6 yr of NICER Data}",
    eprint = "2406.14466",
    archivePrefix = "arXiv",
    primaryClass = "astro-ph.HE",
    doi = "10.3847/1538-4357/ad5f1f",
    journal = "Astrophys. J.",
    volume = "974",
    number = "2",
    pages = "294",
    year = "2024"
}

@article{Choudhury:2024xbk,
    author = "Choudhury, Devarshi and others",
    title = "{A NICER View of the Nearest and Brightest Millisecond Pulsar: PSR J0437{\textendash}4715}",
    eprint = "2407.06789",
    archivePrefix = "arXiv",
    primaryClass = "astro-ph.HE",
    doi = "10.3847/2041-8213/ad5a6f",
    journal = "Astrophys. J. Lett.",
    volume = "971",
    number = "1",
    pages = "L20",
    year = "2024"
}

@article{Mauviard:2025dmd,
    author = "Mauviard, Lucien and others",
    title = "{A NICER View of the 1.4 M$_{\odot}$ Edge-on Pulsar PSR J0614-3329}",
    eprint = "2506.14883",
    archivePrefix = "arXiv",
    primaryClass = "astro-ph.HE",
    doi = "10.3847/1538-4357/ae145d",
    journal = "Astrophys. J.",
    volume = "995",
    number = "1",
    pages = "60",
    year = "2025"
}

@article{Fan:2023spm,
    author = "Fan, Yi-Zhong and Han, Ming-Zhe and Jiang, Jin-Liang and Shao, Dong-Sheng and Tang, Shao-Peng",
    title = "{Maximum gravitational mass MTOV=2.25-0.07+0.08M{\ensuremath{\odot}} inferred at about 3{\%} precision with multimessenger data of neutron stars}",
    eprint = "2309.12644",
    archivePrefix = "arXiv",
    primaryClass = "astro-ph.HE",
    doi = "10.1103/PhysRevD.109.043052",
    journal = "Phys. Rev. D",
    volume = "109",
    number = "4",
    pages = "043052",
    year = "2024"
}

@article{Landry:2018dwm,
    author = "Landry, Philippe and Essick, Reed",
    title = "{Nonparametric inference of the neutron star equation of state from gravitational wave observations}",
    doi = "10.1103/PhysRevD.99.084049",
    journal = "Phys. Rev. D",
    volume = "99",
    number = "8",
    pages = "084049",
    year = "2019",
    eprint = "1803.07580",
    archivePrefix = "arXiv",
    primaryClass = "gr-qc"
}

@article{Essick:2019gia,
    author = "Essick, Reed and Landry, Philippe and Holz, Daniel E.",
    title = "{Nonparametric inference of neutron star composition}",
    doi = "10.1103/PhysRevD.101.063007",
    journal = "Phys. Rev. D",
    volume = "101",
    number = "6",
    pages = "063007",
    year = "2020",
    eprint = "1910.09740",
    archivePrefix = "arXiv",
    primaryClass = "astro-ph.HE"
}

@article{Skilling2006,
    author = "Skilling, John",
    title = "{Nested sampling for general Bayesian computation}",
    doi = "10.1214/06-BA127",
    journal = "Bayesian Analysis",
    volume = "1",
    number = "4",
    pages = "833-859",
    year = "2006"
}

@book{Jolliffe2002,
    author = "Jolliffe, Ian T.",
    title = "{Principal Component Analysis}",
    edition = "2nd",
    publisher = "Springer",
    address = "New York",
    year = "2002"
}

\end{document}